\documentclass[prl,superscriptaddress,twocolumn,notitlepage]{revtex4-1}
\usepackage[latin9]{inputenc}
\usepackage{color}
\usepackage{amsmath}
\usepackage{amssymb}
\usepackage{graphicx}
\usepackage[unicode=true,
 bookmarks=true,bookmarksnumbered=false,bookmarksopen=false,
 breaklinks=false,pdfborder={0 0 1},backref=false,colorlinks=true]
 {hyperref}
\hypersetup{
 linkcolor=magenta,urlcolor=blue,citecolor=blue,pdfstartview={FitH},urlcolor=blue}

\makeatletter
\usepackage{amsmath}
\usepackage{amssymb}
\usepackage{graphicx}
\usepackage{graphbox}

\usepackage{amsfonts}
\usepackage{tabularx}
\usepackage{dcolumn}
\usepackage{bm}
\usepackage{epstopdf}
\usepackage{times}

\newcommand{\ket}[1]{|#1\rangle}
\newcommand{\bra}[1]{\langle #1 |}
\newcommand{\innerproduct}[2]{\langle #1 | #2 \rangle}
\newcommand{\avr}[1]{\langle #1 \rangle}
\newcommand{\tr}[1]{\textnormal{Tr}(#1)}

\makeatother

\begin{document}
\title{Solving the Liouvillian Gap with Artificial Neural Networks}
\author{Dong Yuan}
\thanks{These authors contributed equally to this work.}
\affiliation{Center for Quantum Information, IIIS, Tsinghua University, Beijing
100084, People's Republic of China}
\affiliation{Department of Physics, Tsinghua University, Beijing 100084, People's
Republic of China}
\author{He-Ran Wang}
\thanks{These authors contributed equally to this work.}
\affiliation{Department of Physics, Tsinghua University, Beijing 100084, People's
Republic of China}
\affiliation{Institute for Advanced Study, Tsinghua University, Beijing 100084,
People's Republic of China}
\author{Zhong Wang}
\email{wangzhongemail@tsinghua.edu.cn}

\affiliation{Institute for Advanced Study, Tsinghua University, Beijing 100084,
People's Republic of China}

\author{Dong-Ling Deng}
\email{dldeng@tsinghua.edu.cn}
\affiliation{Center for Quantum Information, IIIS, Tsinghua University, Beijing
100084, People's Republic of China}
\affiliation{Shanghai Qi Zhi Institute, 41st Floor, AI Tower, No. 701 Yunjin Road, Xuhui District, Shanghai 200232, China}

\begin{abstract}
We propose a machine-learning inspired variational method to obtain
the Liouvillian gap, which plays a crucial role in characterizing
the relaxation time and dissipative phase transitions of open quantum
systems. By using the ``spin bi-base mapping'', we map the density
matrix to a pure restricted-Boltzmann-machine (RBM) state and transform
the Liouvillian superoperator to a rank-two non-Hermitian operator.
The Liouvillian gap can be obtained by a variational real-time evolution
algorithm under this non-Hermitian operator. We apply our method to
the dissipative Heisenberg model in both one and two dimensions.
For the isotropic case, we find that the Liouvillian gap can be analytically
obtained and in one dimension even the whole Liouvillian spectrum
can be exactly solved using the Bethe ansatz method. By
comparing our numerical results with their analytical counterparts,
we show that the Liouvillian gap could be accessed by the RBM approach
efficiently to a desirable accuracy, regardless of the dimensionality
and entanglement properties.
\end{abstract}

\maketitle

\begin{figure}
\hspace*{-0.45\textwidth}
\includegraphics[width=0.45\textwidth]{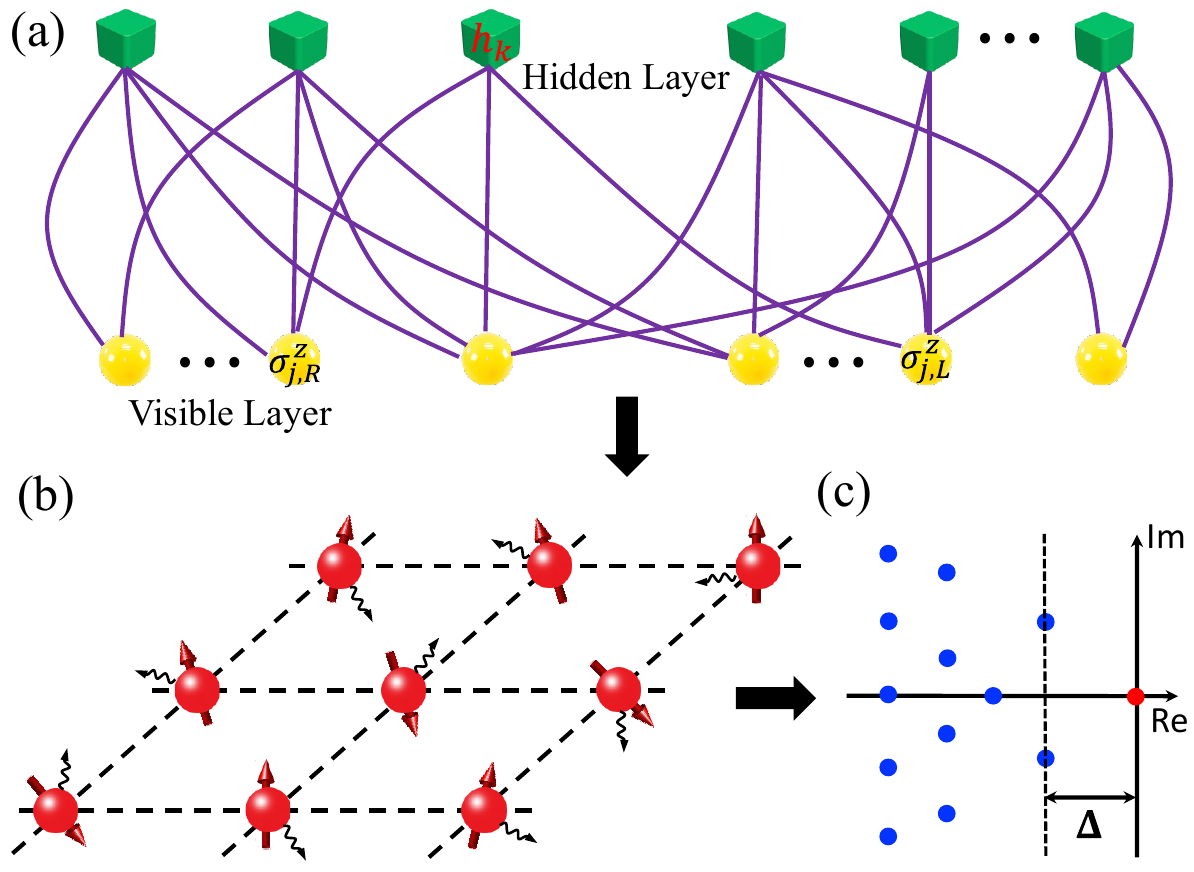} 
\caption{A sketch of the essential idea for accessing the Liouvillian gap with
artificial neural networks. (a) The restricted Boltzmann machine (RBM) representation of quantum many-body density matrices under the ``spin bi-base mapping".  (b) The dissipative spin-$1/2$ XYZ model defined on a 2D square lattice.  
(c) A pictorial sketch of the Liouvillian spectrum. The red dot corresponds to the non-equilibrium steady state with zero Liouvillian eigenvalue, whereas the blue dots on the vertical dotted line correspond to the first decay modes.  $\Delta$ denotes the Liouvillian gap. 
}
\label{Illustration of the idea} 
\end{figure}

Studies of open quantum systems have attracted tremendous attention
across a wide variety of fields \cite{breuer2002openquanttheory,lidar2019lecture},
ranging from condensed matter physics to quantum simulation \cite{noh2016quantum}
and quantum information processing \cite{Nielsen2010quantum}. Within
the Markovian approximation, the dynamics of an open quantum system
is governed by the Lindblad master equation. Relevant to this equation,
a fundamental quantity that characterizes the relaxation time and
dissipative phase transitions of open quantum systems is the 
Liouvillian gap, defined as the gap between the first and second largest
real parts of the eigenspectrum of the Liouvillian superoperator.
For quantum many-body systems, obtaining the Liouvillian gap poses
a notorious challenge for both analytical and numerical approaches,
owing to the exponential scaling of the Hilbert space dimension with
the system size. Despite a few pronounced solvable examples \cite{Medvedyeva2016Exact,prosen2008third,Banchi2017Driven,Rowlands2018Noisy,Ribeiro2019Integrable,Shibata2019Dissipative,Shibata2019ExactsolIsingchain,Lucas2020spectralrandomLiouvillian,ziolkowska2020yang,nakagawa2020exact,buvca2020bethe},
a flexible and scalable numerical approach to compute the Liouvillian
gap is still lacking hitherto. Here, we add this crucial yet missing
block by introducing a generic machine-learning inspired variational
method, with a focus on the restricted-Boltzmann-machine (RBM) architecture
(see Fig. \ref{Illustration of the idea}).

From the numerical computation point of view, computing the Liouvillian
gap is a formidable task in general. In fact, it has been rigorously
proved that the general spectral gap problem is undecidable even for
closed quantum systems \cite{Lloyd1993Quantum,Cubitt2015Undecidability,Bausch2020Undecidability}:
there exists \textit{no} algorithm to determine whether an arbitrary
Hamiltonian is gapped or not. Computing the Liouvillian gap is even
harder in general. Fortunately, physical Hamiltonians or Liouvillians
of practical interest often bear special structures, which may enable
them to circumvent the undecidability and make their spectral gaps
accessible by certain numerical methods.  In particular, for solving Lindblad master equations to obtain the dynamics and non-equilibrium steady states of open quantum systems, a number of notable algorithms have been proposed \cite{weimer2019simulationRMP}, including matrix-product state and tensor network approaches \cite{Verstraete2004Matrix,Zwolak2004Mixed,ZiCai2013AlgebraicversusExponential,Mascarenhas2015Matrix, Cui2015Variational,Marko2015Lgapscaling,Werner2016Positive,Gangat2017Steady,kshetrimayum2017simpleTNsteady},  corner-space renormalization \cite{FinazziCorner,Rota2019Quantum}, cluster mean-field \cite{Jin2016Cluster}, and quantum Monte Carlo \cite{Nagy2018Driven,Yan2018Interacting, Casteels2018Gutzwiller}.   More recently, machine learning \cite{goodfellow2016deep}  approaches based on artificial neural networks have also been invoked to tackle this problem \cite{Hartmann2019Neural,Vicentini2019Variational, Nagy2019Variational,Yoshioka2019Constructing}. The essential idea is to use neural network quantum states \cite{Carleo2016Solving,Torlai2018Latent,Carrasquilla2019Reconstructing,Banchi2018Modelling,Cai2018Approximating,Markus2020manybodydynamics2D}, especially the RBM states \cite{Carleo2016Solving,Torlai2018Latent}, to serve as ansatz density matrices for open quantum systems and adopt the stochastic reconfiguration (SR) method \cite{Sorella2007Weak} to obtain their dynamics and steady states by  solving the master equation variationally. Owing to the structure flexibility and long-range connections of neural networks, such approaches admit the striking merit of generic applicability to high dimensional systems with even volume-law entanglement \cite{Deng2017Quantum}.

In this paper, we introduce a variational method based on the RBM representation to compute the Liouvillian gap.  We use the ``spin bi-base mapping" to map the density matrix to a pure state that is conveniently represented by RBMs. Under this mapping, the Liouvillian superoperator reduces to a rank-two non-Hermitian operator and the Liouvillian gap can be computed by the variational SR algorithm \cite{Sorella2007Weak}. To demonstrate and benchmark the accuracy and
efficiency, we apply our method to the dissipative XYZ (also known as Heisenberg) models in both one and two dimensions (2D). We find that for the isotropic case (the dissipative XXZ model), the Liouvillian gap is always equal to half of the dissipation rate, independent of the coupling strengths, system sizes, and lattice geometry. Inspired by this observation, we show that for the dissipative XXZ model  the Liouvillian gap is indeed exactly solvable, although the Liouvillian spectrum is \textit{not} solvable in general. In 1D, we show that the whole Liouvillian spectrum can be exactly solved using the Bethe ansatz method. These analytic results may be of independent interest. For the anisotropic case, we compare our RBM results with the results from exact diagonalization (ED) and find that they match within a desirable accuracy.

\textit{The Liouvillian gap and RBM approach}.---We consider the following Lindblad master equation \cite{breuer2002openquanttheory,lidar2019lecture}:
\begin{equation}
\frac{\mathrm{d}\rho}{\mathrm{d}t}=-i[H,\rho]+\sum_{\mu}(2L_{\mu}\rho L_{\mu}^{\dagger}-\{L_{\mu}^{\dagger}L_{\mu},\rho\})\equiv\mathcal{L}\rho,
\end{equation}
where $\rho$ denotes the density matrix, $H$ is the Hamiltonian governing the unitary part of the dynamics, 
$L_\mu$ are the jump operators describing the dissipative process, the curly bracket represents the anticommutator, and $\mathcal{L}$ is the Liouvillian superoperator. In general, the index $\mu$ runs over all dissipation channels. For simplicity and concreteness, here we  focus on the case where each lattice site has only  one dissipation channel  and $\mu$ just labels the lattice site. The generalization to the cases with multiple dissipation channels is straightforward. 

The full spectrum of the Liouvillian superoperator $\mathcal{L}$ can be determined by solving the eigenequation: $\mathcal{L}\rho_{k}=\lambda_{k}\rho_{k}$, where $\lambda_k$ is the eigenvalue and $\rho_k$ denotes its corresponding eigenmatrix. Unlike the case of closed quantum systems, the eigenvalues of $\mathcal{L}$ are usually complex and their corresponding eigenmatrices are not necessarily physical (i.e., being Hermitian, trace-one, and semi-positive definite). Moreover, it can be proved \cite{breuer2002openquanttheory} that $\text{Re}(\lambda_k)\leq 0$ for all $k$. For convenience, we sort the eigenvalues by their real parts in decreasing order [$\text{Re}(\lambda_0)\geq \text{Re}(\lambda_1) \geq \cdots $] with the steady state corresponding to $\lambda_0=0$ (for simplicity, we only focus on the case that the steady state is unique and $\lambda_0$ has no degeneracy).  With this convention, the Liouvillian gap is  defined as
\begin{eqnarray}
\Delta=-\text{Re}(\lambda_1).
\end{eqnarray}
The Liouvillian gap is a central and fundamental physical quantity in studying open quantum systems. It determines the relaxation time  from an arbitrary initial state to the steady state and plays a crucial role in characterizing dissipative phase transitions \cite{Rossatto2016Relaxationtime,Kessler2012Dissipative,Minganti2018Spectral} and the exotic chiral damping phenomenon \cite{Song2019NonHermitian}. Yet, for quantum many-body systems the dimension of the Liouville space scales double exponentially  with the system size, rendering the computation of Liouvillian gap notably challenging.  

We employ a spin bi-base mapping, which is also called the Choi-Jamio\l kowski isomorphism \cite{Zwolak2004Mixed,kshetrimayum2017simpleTNsteady}, to map a density matrix to a vector in the computational bases:
\begin{equation}
\rho=\sum_{m,n}\rho_{mn}\ket{m}\bra{n}\quad\Leftrightarrow\quad\tilde{\rho}=\sum_{m,n}\rho_{mn}\ket{m}\otimes\ket{n}. \label{Bi-base mapping}
\end{equation}
Under this mapping, the Liouvillian superoperator $\mathcal{L}$, originally a rank-four tensor, reduces to a rank-two operator $\tilde{\mathcal{L}}=-iH\otimes I+iI\otimes H^{T}+\sum_{\mu}(2L_{\mu}\otimes L_{\mu}^{*}-L_{\mu}^{\dagger}L_{\mu}\otimes I-I\otimes L_{\mu}^{T}L_{\mu}^{*})$, where $I$ denotes the identity matrix and $T$ means the matrix transpose. We consider an open quantum system with $N$ qubits and use a RBM to describe $\tilde{\rho}$ with \cite{Carleo2016Solving}
\begin{eqnarray}
(\rho_{\text{RBM}})_{mn}=\exp{\left[\sum_{j=1}^{N}(a_{j}\sigma_{j,R}^{z}+b_{j}\sigma_{j,L}^{z})\right]}\prod_{k=1}^{M}X_{k}, \label{RBMState}
\end{eqnarray}
where $\sigma^z_{j,R(L)}=\pm 1$ denotes the visible neurons responsible for the $|m\rangle =|\sigma^z_{1,R},\cdots,\sigma^z_{N,R}\rangle$ $\left(|n\rangle =|\sigma^z_{1,L},\cdots,\sigma^z_{N,L}\rangle\right)$ part of $\tilde{\rho}$. $M$ is the number of hidden neurons, and  $X_{k}=\cosh\left(c_{k}+\sum_j W^R_{k,j}\sigma_{j,R}^{z}+\sum_j W^L_{k,j}\sigma_{j,L}^{z}\right)$. Here, $\{a_{j},b_{j},c_{k}\}$ are on-site weight parameters, and $\{W^R_{k,j}$, $W^L_{k,j}\}$ are connection parameters between visible and hidden neurons. We note that the matrix $\rho$ corresponding to $\tilde{\rho}$ may \textit{not} be physical, in contrast to the representations of density matrices introduced in Refs. \cite{Torlai2018Latent,Hartmann2019Neural,Vicentini2019Variational, Nagy2019Variational,Yoshioka2019Constructing}. This is not a problem for our purpose because here we mainly focus on the eigenmatrices of $\mathcal{L}$, which are not physical in general. 

\begin{figure*}
\hspace*{-0.9\textwidth}
\includegraphics[width=0.9\textwidth]{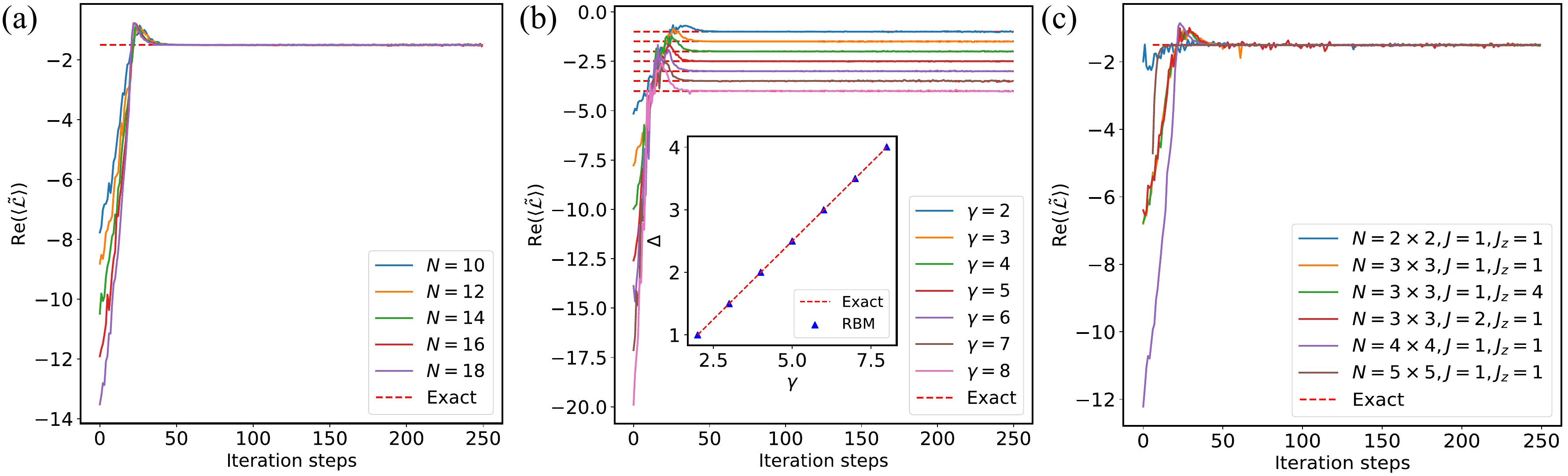}

\caption{Numerical results for the dissipative XXZ model in one [(a) and (b)] and two [(c)] dimensions.  (a) The real part of the expectation value $\text{Re}(\langle\tilde{\mathcal{L}}\rangle)=\text{Re}(\langle \rho'|\tilde{\mathcal{L}}|\rho'\rangle/\langle \rho'|\rho'\rangle)$ as a function of the iteration steps for different lattice sizes $N$. The parameters are chosen as $J_x=J_y=1$, $J_z=2$, and $\gamma=3$. The red dashed horizontal line indicates the exact value of the Liouvillian gap $\Delta$ obtained by exact diagonalization, which can also be derived analytically (see the main text). (b) $\text{Re}(\langle\tilde{\mathcal{L}}\rangle)$  as a function of iteration steps for different dissipation rates.  Here, $J_x=J_y=1$, $J_z=2$, and  $N=10$. The inset shows the linear dependence of the Liouvillian gap obtained by the RBM method on the dissipation rate $\gamma$. (c) $\text{Re}(\langle\hat{\mathcal{L}}\rangle)$   for 2D dissipative XXZ model with $\gamma=3$, and varying lattice sizes and coupling strengths (see \cite{LGviaRBMSuppM} for details). 
\label{fig:NumXXZ}}
\end{figure*}

Given the RBM parametrization of $\tilde{\rho}$, we can now recast the problem of computing the Liouvillian gap as a variational optimization problem  in a subspace orthogonal to the steady state $\rho_0$. Yet, the subtraction of $\rho_0$ is tricky. Analogous to the closed system case, one may regard the first decay modes as the first ``excited states" of $\mathcal{L}$ and then the Liouvillian gap is just the ``first excited energy". Consequently, a possible way to obtain the Liouvillian gap is by first computing the steady state of $\mathcal{L}$ and then appropriately extending the protocols for calculating the excited states of closed systems \cite{Vieijra2020Restricted,Choo2018Symmetries} to open systems. This approach is straightforward, yet technically cumbersome. By noticing the fact that $\mathcal{L^\dagger}I = 0$, $\text{Tr}(\mathcal{L}\rho_k)=0=\lambda_k\text{Tr}(\rho_k)$ and hence $\text{Tr}(\rho_k)=0$ for all $\lambda_k\ne 0$,  a much simpler approach is to construct a new variational matrix $\rho'$ with vanishing trace 
\begin{eqnarray}
\rho'=\alpha\rho'_0+\rho_{\text{RBM}},
\end{eqnarray}
where $\alpha=-\frac{\text{Tr}(\rho_{\text{RBM}})}{\text{Tr}(\rho'_0)}$, $\rho'_0$ is a density matrix with nonzero trace which is not necessary the true steady state, and $\rho_{\text{RBM}}$ is a RBM ansatz state as defined in Eq. (\ref{RBMState}). Since $\text{Tr}(\rho')=0$, $\rho'$ lives in the subspace orthogonal to $\rho_0$. We adapt the SR method \cite{Sorella2007Weak}  to generate the real time evolution of $\rho'$  \cite{LGviaRBMSuppM}. Unlike the case for closed systems, $\tilde{\mathcal{L}}$ is not Hermitian and its right eigenvectors are not orthogonal in general. This non-Hermiticity makes the problem more complicated. Three different cases arise based on the properties of the first decay modes: (i) there is only one first decay mode, then $\rho'\rightarrow \rho_1$ after long enough real-time evolution and the Liouvillian gap can be obtained by $\Delta=-\text{Re}(\langle \rho'|\mathcal{L}|\rho'\rangle/\langle \rho'|\rho'\rangle)$; (ii) there are multiple first decay modes but they are orthogonal to each other, then $\rho'$ will converge to a superposition of these decay modes and $\Delta$ can still be obtained in the same way as in the first case; (iii) there exist multiple first decay modes [denoted as $\rho_1^{(1)}, \rho_1^{(2)},\rho_1^{(3)},\cdots$] which are not orthogonal. In this case, $\rho'$ will converge to a superposition of these decay modes:
$\rho'\rightarrow a_1 \rho_1^{(1)}+a_2 \rho_1^{(2)}+a_3 \rho_1^{(3)}+\cdots$,
where $a_1,a_2,a_3,\cdots$ are coefficients whose values depend on the initialization of $\rho'$. Due to the non-orthogonality, $\langle \rho_1^{(i)}|\mathcal{L}|\rho_1^{(j)}\rangle\neq 0$ for $i\neq j$ and $\Delta\neq -\text{Re}(\langle \rho'|\mathcal{L}|\rho'\rangle/\langle \rho'|\rho'\rangle)$ at this stage in general. To overcome this  problem, we should add another imaginary time evolution for $\rho'$ under $i\mathcal{L}$ with a smaller learning rate so that it converges further to the first decay mode with the minimal imaginary part. After this modification, $\Delta$ can be obtained again by computing $-\text{Re}(\langle \rho'|\mathcal{L}|\rho'\rangle/\langle \rho'|\rho'\rangle)$ \cite{LGviaRBMSuppM}.
We stress the fact that the non-Hermiticity of $\mathcal{L}$ makes the computation of the Liouvillian gap much subtler than computing the energy gap for a Hermitian Hamiltonian, as discussed above.  However, the computational complexity of our RBM approach does not increase too much and is still favorable \cite{LGviaRBMSuppM}. 

\textit{Concrete examples.}---We consider the dissipative spin-1/2 XYZ model, where each spin has a Heisenberg type interaction with its nearest neighboring spins and is subject to a dissipation process into the $|S^z=-1/2\rangle$ state: 
\begin{eqnarray}
\frac{d \rho}{d t}=-i[H,\rho]+\frac{\gamma}{2}\sum_j [2S_j^-\rho S_j^+-\{S_j^+S_j^-,\rho\}].\label{XYZmodel} 
\end{eqnarray}
Here $H =\sum_{\langle j,k\rangle}\left(J_{x}S_{j}^{x}S_{k}^{x}+J_{y}S_{j}^{y}S_{k}^{y}+J_{z}S_{j}^{z}S_{k}^{z}\right)$, $S^{\mu}=\frac{1}{2}\sigma^{\mu}$ with $\sigma^{\mu}$ being the Pauli matrix ($\mu=x,y,z$), $J_{\mu}$ denotes the coupling constant between nearest neighboring spins, and $S^{\pm}=S^x\pm i S^y$. The dynamics and steady state properties of this dissipative XYZ model have already been widely studied \cite{Jin2016Cluster,Lee2013Unconventional,Rota2017Critical,Rota2018Dynamical,Casteels2018Gutzwiller,
Huybrechts2019Cluster,Hartmann2019Neural, Nagy2019Variational}. In particular, in 2D it exhibits a dissipative phase transition between a paramagnetic and a ferromagnetic phase \cite{Jin2016Cluster,Lee2013Unconventional,Rota2017Critical,Rota2018Dynamical,Casteels2018Gutzwiller,Huybrechts2019Cluster}, which originates from the competition between the unitary evolution and the incoherent dynamics. Different from the previous works that mainly focus on dynamics or steady state properties, here we focus on the computation of  the Liouvillian gap instead. 

We apply the introduced RBM approach to compute the Liouvillian gap for the dissipative XYZ model in both one and two dimensions. For the isotropic case $J_x=J_y=J$, the XYZ model reduces to the XXZ model and our numerical results are shown in Fig. \ref{fig:NumXXZ}. From this figure, it is evident that the real part of the expectation value $\text{Re}(\langle\tilde{\mathcal{L}}\rangle)=\text{Re}(\langle \rho'|\tilde{\mathcal{L}}|\rho'\rangle/\langle \rho'|\rho'\rangle)$ converges quickly to the exact value of the Liouvillian gap, validating the effectiveness of the RBM method. To measure the accuracy, we define the relative error $\epsilon_{\text{rel}}=|(\Delta_{\text{RBM}}-\Delta_{\text{Ex}})/\Delta_{\text{Ex}}|$ and find that $\epsilon_{\text{rel}}\approx 10^{-2}$ after around $50$ iteration steps for all the scenarios shown in Fig. \ref{fig:NumXXZ}. We mention that the accuracy can be systematically improved by increasing the number of hidden neurons or the length of the Markov chain  used in the SR algorithm \cite{LGviaRBMSuppM}. 

An interesting observation from our RBM results shown in Fig. \ref{fig:NumXXZ} is that for the dissipative XXZ model, the Liouvillian gap is always equal to half of the dissipation rate $\Delta=\gamma/2$, independent of the coupling strengths $J$ and $J_z$, system size, boundary condition, and the model dimensionality. This inspired us to suspect that the Liouvillian gap is exactly solvable owing to special structures of the dissipative XXZ model. Indeed, we find that $\Delta$ can be analytically derived as below. After the spin bi-base mapping, the Liouvillian superoperator $\mathcal{L}$ for the dissipative XXZ model is mapped to: $\tilde{\mathcal{L}}= -i\sum_{\langle i,j\rangle}[J_{z}(S_{i,R}^{z}S_{j,R}^{z}-S_{i,L}^{z}S_{j,L}^{z})
+\frac{1}{2}J(S_{i,R}^{+}S_{j,R}^{-}+S_{i,R}^{-}S_{j,R}^{+}-S_{i,L}^{+}S_{j,L}^{-}-S_{i,L}^{-}S_{j,L}^{+})]
+\frac{\gamma}{2}\sum_{i}(2S_{i,R}^{-}S_{i,L}^{-}-S_{i,R}^{z}-S_{i,L}^{z}-1)$.
One may regard $\tilde{\mathcal{L}}$ as a non-Hermitian Hamiltonian for two copies of the original dissipative system, with one copy corresponding to the right ket (the $|m\rangle$ part) and the other copy corresponding to the left bra (the $|n\rangle$ part) in Eq. (\ref{Bi-base mapping}). We then rearrange the terms in $\tilde{\mathcal{L}}$ accordingly: 
\begin{eqnarray}
\tilde{\mathcal{L}}=H_R+H_L+\gamma \sum_i D_i,
\end{eqnarray}
where $D_i=S_{i,R}^{-}S_{i,L}^{-}$ describes the couplings between the left and right spins, and $H_{R}=-i\sum_{\langle i, j\rangle} [J(S_{i,R}^{+}S_{j,R}^{-}+S_{i,R}^{-}S_{j,R}^{+})+J_z S_{i,R}^{z}S_{j,R}^{z}]-\frac{\gamma}{2}\sum_i (S^z_{i,R}+\frac{1}{2})$  and $H_{L}=i\sum_{\langle i, j\rangle} [J(S_{i,L}^{+}S_{j,L}^{-}+S_{i,L}^{-}S_{j,L}^{+})+J_z S_{i,L}^{z}S_{j,L}^{z}]-\frac{\gamma}{2}\sum_i (S^z_{i,L}+\frac{1}{2})$ denote the Hamiltonians for the right and left subsystems, respectively. It is easy to observe that $[H_L, H_R]=0$ since they belong to different subsystems, and both $H_L$ and $H_R$ have a $U(1)$ symmetry, namely their total $S^z$ is conserved respectively. Consequently, $\tilde{\mathcal{L}'}=H_R+H_L$ can be block-diagonalized with each block maintaining a fixed total $S^z$. Following Ref. \cite{torre2014closedform}, in the bases where $\tilde{\mathcal{L}'}$ is block-diagonal, each term $D_i$ is just an upper triangular matrix with vanishing diagonal terms, thus adding these terms will only alter the eigenstates but \textit{not} the eigenvalues of $\tilde{\mathcal{L}'}$.    The eigenspectrum of $\tilde{\mathcal{L}}$ is exactly the same as $\tilde{\mathcal{L}'}$ \cite{LGviaRBMSuppM}. In addition, it is easy to observe that the steady state of $\mathcal{L}$ is the state with all spins pointing down due to the dissipation process. Noting that $\tilde{\mathcal{L}}'$ contains only imaginary XXZ interactions and a magnetic field with strength $\frac{\gamma}{2}$, hence the real part of the spectrum of $\tilde{\mathcal{L}}$ is just $-\frac{\gamma}{2}m$, where $m$ denotes the number of ``magnons" created from the steady state (the number of spins flipped from down to up). As a result, the desired Liouvillian gap corresponds to a single-magnon excitation, which leads to $\Delta=\frac{\gamma}{2}$, independent of $J,J_z$, system size, and lattice geometry. In 1D with periodic boundary condition, the whole Liouvillian spectrum can be deduced from the Bethe ansatz solution \cite{LGviaRBMSuppM}:
\begin{equation}
E(\{k_{j}\}_{m})=-\frac{\gamma}{2}m\pm i\sum_{j=1}^{m}(2J\textnormal{cos}k_{j}-J_{z}). \label{spectrum}
\end{equation}

\begin{figure}
\hspace*{-0.47\textwidth}
\centering 
\includegraphics[width=0.47\textwidth]{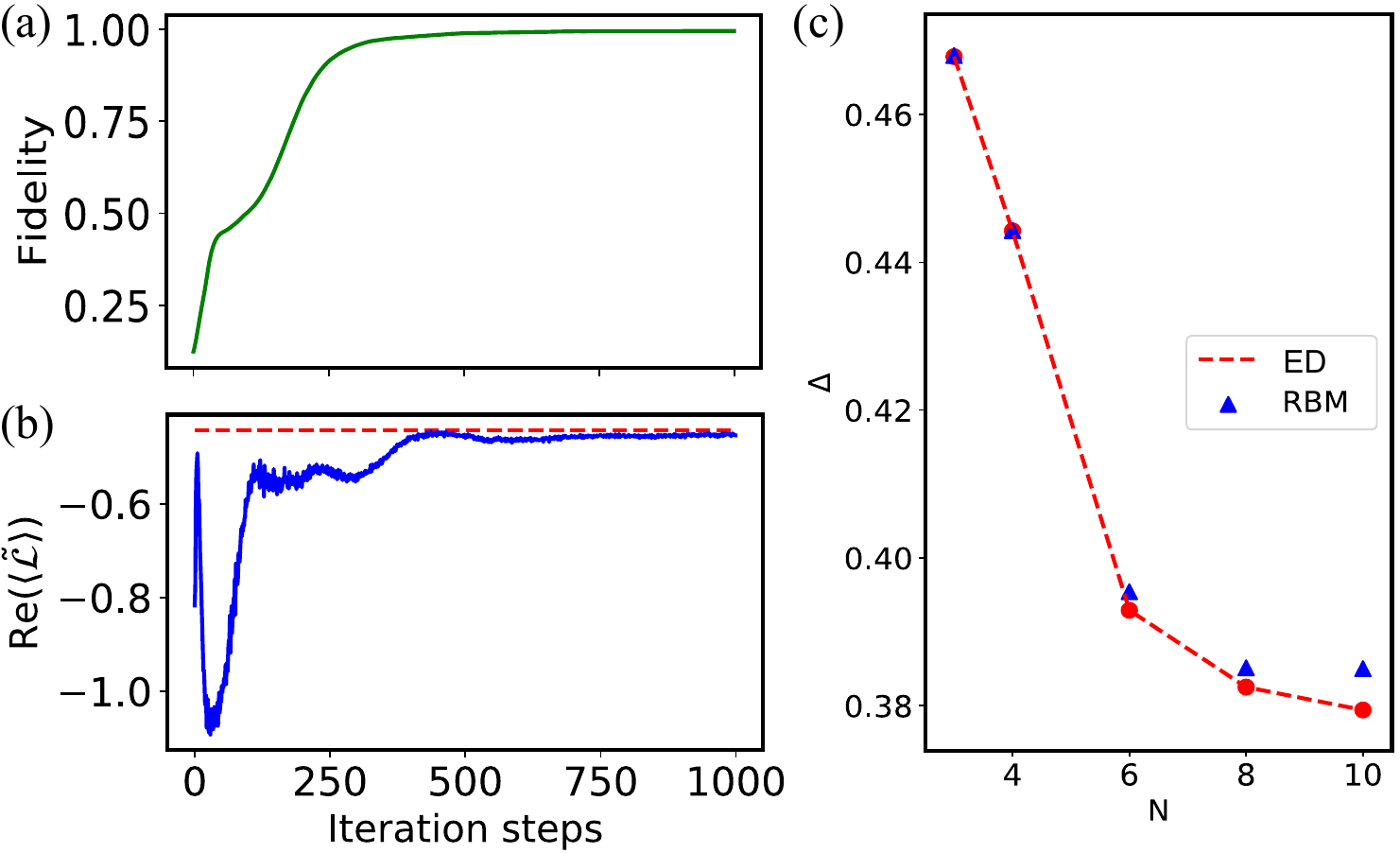} 
\caption{Numerical results for the dissipative XYZ model in 1D. (a) and (b) show respectively the fidelity between $|\rho'\rangle$ and $|\rho_1\rangle$, and the expectation value $\text{Re}(\langle \tilde{\mathcal{L}}\rangle)$, as functions of the iteration steps for $N=4$. After around 500 iterations, $\text{Re}(\langle \tilde{\mathcal{L}}\rangle)$ converges to the exact value of the Liouvillian gap $\Delta$ [indicated by the red dashed line in (b)] with relative error $\epsilon_{\text{rel}}\approx 10^{-2}$ and the fidelity approaches one ($|\langle \rho'| \rho_1\rangle|\approx 0.991$). (c) The comparison between the RBM and ED results with varying system sizes.   Here, the model parameters are chosen as $J_x=4$, $J_y=0.5$, $J_z=2$, and $\gamma=1$ \cite{LGviaRBMSuppM}. 
} \label{figure: XYZresults}
\end{figure}

 We now turn to the anisotropic case with $J_x\neq J_y$, where the Liouvillian gap cannot be solved analytically in general. 
 Our numerical results are shown in Fig. \ref{figure: XYZresults}. From this figure, it is clear that  
 our RBM results match the exact results  from exact diagonalization within a reasonable accuracy. We note that, in Fig.  \ref{figure: XYZresults}(c),  the apparent deviation of the RBM result for $N=10$ from its exact value is due to the small plotting range.  A closer examination shows that for this point, the relative error is $\epsilon_{\text{rel}}=1.46\times 10^{-2}$ in fact. In comparison with the case of the XXZ model, we find that the convergence for the XYZ model is notably slower. The reason for this is that for the XXZ model, its multiple first decay modes are orthogonal to each other. Hence, to obtain $\Delta$, $\rho'$ only needs to converge to a subspace spanned by these modes. Whereas for the XYZ model, depending on the parameters there exist either only one or multiple but non-orthogonal first decay modes \cite{LGviaRBMSuppM}. Thus,  as discussed previously, to obtain $\Delta$ accurately in this case, $\rho'$ needs to converge to a single decay mode,  which demands extra iteration steps and possibly more hidden neurons to increase the representation power. We note that our RBM approach may also carry over straightforwardly to systems with higher spins, interacting bosons \cite{Saito2017Solving, Choo2018Symmetries} and fermions \cite{Nomura2017Restricted,Choo2020Fermionic}, or models with long-range interactions \cite{Deng2017Quantum,LGviaRBMSuppM}. 

\textit{Discussion and conclusion.}--- We mention that the general spectral gap problem has been mathematically proved to  be an undecidable problem \cite{Lloyd1993Quantum,Cubitt2015Undecidability,Bausch2020Undecidability} from the computational complexity perspective \cite{Arora2009Computational}, implying that there exists \textit{no} algorithm to determine whether an arbitrary model is gapped or gapless. As a result, we cannot  expect that our RBM approach could solve the Liouvillian gap for all possible Liouvillian superoperators. 
In fact, no algorithm is capable of doing this due to the undecidability of the problem. Finding out the key properties of the Liouvillian superoperators that warrant the effectiveness of the RBM approach is of both fundamental and practical importance. Yet, this may require new physical concepts and a deeper understanding of artificial neural networks, similar to the case of how we understand the effectiveness of the density-matrix-renormalization-group algorithm \cite{Schollwock2005TheDMRG} from the entanglement perspective. 
In addition, one may also use other neural networks, such as deep Boltzmann machine \cite{Gao2017Efficient} or feedforward neural networks \cite{Cai2018Approximating}, to compute the the Liouvillian gap. More recently, a quantum-classical hybrid algorithm based on deep quantum neural networks has also been introduced to solve the steady states and dynamics for open systems \cite{Liu2020Solving}. It would also be interesting to extend this approach to compute the Liouvillian gap through noisy intermediate-scale quantum devices \cite{Preskill2018quantum}.

In summary, we have introduced a machine learning based approach to compute the Liouvillian gap for open quantum systems, which is generally applicable to  high dimensional systems with massive entanglement. The accuracy and effectiveness of this approach have been  benchmarked with numerical examples for the dissipative Heisenberg model in both one and two dimensions. Based on the numerical results, we found that the Liouvillian gap is exactly solvable for the dissipative XXZ model regardless of the system size and lattice geometry. These analytic results are of independent interest as well and may inspire subsequent analytical studies. 

We acknowledge helpful discussions with Fei Song, Shunyao Zhang and computational resources provided by Yiyang Wu and Benda Xu. This work was supported by the start-up fund from Tsinghua University (Grant No. 53330300320), NSFC under Grants No. 11674189 and 12075128, and the Shanghai Qi Zhi Institute.

\textit{Note added.}--- After this work was finished, we became aware of a recent work \cite{buvca2020bethe}, which also analytically studied the Liouvillian spectrum based on the Bethe ansatz method.

\bibliographystyle{apsrev4-1-title}
\bibliography{Dengbib,Dongbib,Heranbib,LGviaRBM,NonHermRefs}

\begin{thebibliography}{72}%
\makeatletter
\providecommand \@ifxundefined [1]{%
 \@ifx{#1\undefined}
}%
\providecommand \@ifnum [1]{%
 \ifnum #1\expandafter \@firstoftwo
 \else \expandafter \@secondoftwo
 \fi
}%
\providecommand \@ifx [1]{%
 \ifx #1\expandafter \@firstoftwo
 \else \expandafter \@secondoftwo
 \fi
}%
\providecommand \natexlab [1]{#1}%
\providecommand \enquote  [1]{``#1''}%
\providecommand \bibnamefont  [1]{#1}%
\providecommand \bibfnamefont [1]{#1}%
\providecommand \citenamefont [1]{#1}%
\providecommand \href@noop [0]{\@secondoftwo}%
\providecommand \href [0]{\begingroup \@sanitize@url \@href}%
\providecommand \@href[1]{\@@startlink{#1}\@@href}%
\providecommand \@@href[1]{\endgroup#1\@@endlink}%
\providecommand \@sanitize@url [0]{\catcode `\\12\catcode `\$12\catcode
  `\&12\catcode `\#12\catcode `\^12\catcode `\_12\catcode `\%12\relax}%
\providecommand \@@startlink[1]{}%
\providecommand \@@endlink[0]{}%
\providecommand \url  [0]{\begingroup\@sanitize@url \@url }%
\providecommand \@url [1]{\endgroup\@href {#1}{\urlprefix }}%
\providecommand \urlprefix  [0]{URL }%
\providecommand \Eprint [0]{\href }%
\providecommand \doibase [0]{http://dx.doi.org/}%
\providecommand \selectlanguage [0]{\@gobble}%
\providecommand \bibinfo  [0]{\@secondoftwo}%
\providecommand \bibfield  [0]{\@secondoftwo}%
\providecommand \translation [1]{[#1]}%
\providecommand \BibitemOpen [0]{}%
\providecommand \bibitemStop [0]{}%
\providecommand \bibitemNoStop [0]{.\EOS\space}%
\providecommand \EOS [0]{\spacefactor3000\relax}%
\providecommand \BibitemShut  [1]{\csname bibitem#1\endcsname}%
\let\auto@bib@innerbib\@empty
\bibitem [{\citenamefont {Breuer}\ and\ \citenamefont
  {Petruccione}(2007)}]{breuer2002openquanttheory}%
  \BibitemOpen
  \bibfield  {author} {\bibinfo {author} {\bibfnamefont {H.-P.}\ \bibnamefont
  {Breuer}}\ and\ \bibinfo {author} {\bibfnamefont {F.}~\bibnamefont
  {Petruccione}},\ }\href@noop {} {\emph {\bibinfo {title} {The theory of open
  quantum systems}}}\ (\bibinfo  {publisher} {Oxford University Press,
  Oxford},\ \bibinfo {year} {2007})\BibitemShut {NoStop}%
\bibitem [{\citenamefont {Lidar}(2019)}]{lidar2019lecture}%
  \BibitemOpen
  \bibfield  {author} {\bibinfo {author} {\bibfnamefont {D.~A.}\ \bibnamefont
  {Lidar}},\ }\bibfield  {title} {\enquote {\bibinfo {title} {Lecture notes on
  the theory of open quantum systems},}\ }\href
  {https://arxiv.org/abs/1902.00967} {\bibfield  {journal} {\bibinfo  {journal}
  {arXiv:1902.00967}\ } (\bibinfo {year} {2019})}\BibitemShut {NoStop}%
\bibitem [{\citenamefont {Noh}\ and\ \citenamefont
  {Angelakis}(2016)}]{noh2016quantum}%
  \BibitemOpen
  \bibfield  {author} {\bibinfo {author} {\bibfnamefont {C.}~\bibnamefont
  {Noh}}\ and\ \bibinfo {author} {\bibfnamefont {D.~G.}\ \bibnamefont
  {Angelakis}},\ }\bibfield  {title} {\enquote {\bibinfo {title} {Quantum
  simulations and many-body physics with light},}\ }\href
  {https://iopscience.iop.org/article/10.1088/0034-4885/80/1/016401} {\bibfield
   {journal} {\bibinfo  {journal} {Rep. Prog. in Phys.}\ }\textbf {\bibinfo
  {volume} {80}},\ \bibinfo {pages} {016401} (\bibinfo {year}
  {2016})}\BibitemShut {NoStop}%
\bibitem [{\citenamefont {Nielsen}\ and\ \citenamefont
  {Chuang}(2010)}]{Nielsen2010quantum}%
  \BibitemOpen
  \bibfield  {author} {\bibinfo {author} {\bibfnamefont {M.~A.}\ \bibnamefont
  {Nielsen}}\ and\ \bibinfo {author} {\bibfnamefont {I.~L.}\ \bibnamefont
  {Chuang}},\ }\href@noop {} {\emph {\bibinfo {title} {Quantum computation and
  quantum information}}}\ (\bibinfo  {publisher} {Cambridge university press},\
  \bibinfo {year} {2010})\BibitemShut {NoStop}%
\bibitem [{\citenamefont {Medvedyeva}\ \emph {et~al.}(2016)\citenamefont
  {Medvedyeva}, \citenamefont {Essler},\ and\ \citenamefont
  {Prosen}}]{Medvedyeva2016Exact}%
  \BibitemOpen
  \bibfield  {author} {\bibinfo {author} {\bibfnamefont {M.~V.}\ \bibnamefont
  {Medvedyeva}}, \bibinfo {author} {\bibfnamefont {F.~H.~L.}\ \bibnamefont
  {Essler}}, \ and\ \bibinfo {author} {\bibfnamefont {T.~c.~v.}\ \bibnamefont
  {Prosen}},\ }\bibfield  {title} {\enquote {\bibinfo {title} {Exact bethe
  ansatz spectrum of a tight-binding chain with dephasing noise},}\ }\href
  {\doibase 10.1103/PhysRevLett.117.137202} {\bibfield  {journal} {\bibinfo
  {journal} {Phys. Rev. Lett.}\ }\textbf {\bibinfo {volume} {117}},\ \bibinfo
  {pages} {137202} (\bibinfo {year} {2016})}\BibitemShut {NoStop}%
\bibitem [{\citenamefont {Prosen}(2008)}]{prosen2008third}%
  \BibitemOpen
  \bibfield  {author} {\bibinfo {author} {\bibfnamefont {T.}~\bibnamefont
  {Prosen}},\ }\bibfield  {title} {\enquote {\bibinfo {title} {Third
  quantization: a general method to solve master equations for quadratic open
  fermi systems},}\ }\href
  {https://iopscience.iop.org/article/10.1088/1367-2630/10/4/043026/meta}
  {\bibfield  {journal} {\bibinfo  {journal} {New Journal of Physics}\ }\textbf
  {\bibinfo {volume} {10}},\ \bibinfo {pages} {043026} (\bibinfo {year}
  {2008})}\BibitemShut {NoStop}%
\bibitem [{\citenamefont {Banchi}\ \emph {et~al.}(2017)\citenamefont {Banchi},
  \citenamefont {Burgarth},\ and\ \citenamefont
  {Kastoryano}}]{Banchi2017Driven}%
  \BibitemOpen
  \bibfield  {author} {\bibinfo {author} {\bibfnamefont {L.}~\bibnamefont
  {Banchi}}, \bibinfo {author} {\bibfnamefont {D.}~\bibnamefont {Burgarth}}, \
  and\ \bibinfo {author} {\bibfnamefont {M.~J.}\ \bibnamefont {Kastoryano}},\
  }\bibfield  {title} {\enquote {\bibinfo {title} {Driven quantum dynamics:
  Will it blend?}}\ }\href {\doibase 10.1103/PhysRevX.7.041015} {\bibfield
  {journal} {\bibinfo  {journal} {Phys. Rev. X}\ }\textbf {\bibinfo {volume}
  {7}},\ \bibinfo {pages} {041015} (\bibinfo {year} {2017})}\BibitemShut
  {NoStop}%
\bibitem [{\citenamefont {Rowlands}\ and\ \citenamefont
  {Lamacraft}(2018)}]{Rowlands2018Noisy}%
  \BibitemOpen
  \bibfield  {author} {\bibinfo {author} {\bibfnamefont {D.~A.}\ \bibnamefont
  {Rowlands}}\ and\ \bibinfo {author} {\bibfnamefont {A.}~\bibnamefont
  {Lamacraft}},\ }\bibfield  {title} {\enquote {\bibinfo {title} {Noisy spins
  and the richardson-gaudin model},}\ }\href {\doibase
  10.1103/PhysRevLett.120.090401} {\bibfield  {journal} {\bibinfo  {journal}
  {Phys. Rev. Lett.}\ }\textbf {\bibinfo {volume} {120}},\ \bibinfo {pages}
  {090401} (\bibinfo {year} {2018})}\BibitemShut {NoStop}%
\bibitem [{\citenamefont {Ribeiro}\ and\ \citenamefont
  {Prosen}(2019)}]{Ribeiro2019Integrable}%
  \BibitemOpen
  \bibfield  {author} {\bibinfo {author} {\bibfnamefont {P.}~\bibnamefont
  {Ribeiro}}\ and\ \bibinfo {author} {\bibfnamefont {T.~c.~v.}\ \bibnamefont
  {Prosen}},\ }\bibfield  {title} {\enquote {\bibinfo {title} {Integrable
  quantum dynamics of open collective spin models},}\ }\href {\doibase
  10.1103/PhysRevLett.122.010401} {\bibfield  {journal} {\bibinfo  {journal}
  {Phys. Rev. Lett.}\ }\textbf {\bibinfo {volume} {122}},\ \bibinfo {pages}
  {010401} (\bibinfo {year} {2019})}\BibitemShut {NoStop}%
\bibitem [{\citenamefont {Shibata}\ and\ \citenamefont
  {Katsura}(2019{\natexlab{a}})}]{Shibata2019Dissipative}%
  \BibitemOpen
  \bibfield  {author} {\bibinfo {author} {\bibfnamefont {N.}~\bibnamefont
  {Shibata}}\ and\ \bibinfo {author} {\bibfnamefont {H.}~\bibnamefont
  {Katsura}},\ }\bibfield  {title} {\enquote {\bibinfo {title} {Dissipative
  spin chain as a non-hermitian kitaev ladder},}\ }\href {\doibase
  10.1103/PhysRevB.99.174303} {\bibfield  {journal} {\bibinfo  {journal} {Phys.
  Rev. B}\ }\textbf {\bibinfo {volume} {99}},\ \bibinfo {pages} {174303}
  (\bibinfo {year} {2019}{\natexlab{a}})}\BibitemShut {NoStop}%
\bibitem [{\citenamefont {Shibata}\ and\ \citenamefont
  {Katsura}(2019{\natexlab{b}})}]{Shibata2019ExactsolIsingchain}%
  \BibitemOpen
  \bibfield  {author} {\bibinfo {author} {\bibfnamefont {N.}~\bibnamefont
  {Shibata}}\ and\ \bibinfo {author} {\bibfnamefont {H.}~\bibnamefont
  {Katsura}},\ }\bibfield  {title} {\enquote {\bibinfo {title} {Dissipative
  quantum ising chain as a non-hermitian ashkin-teller model},}\ }\href
  {\doibase 10.1103/PhysRevB.99.224432} {\bibfield  {journal} {\bibinfo
  {journal} {Phys. Rev. B}\ }\textbf {\bibinfo {volume} {99}},\ \bibinfo
  {pages} {224432} (\bibinfo {year} {2019}{\natexlab{b}})}\BibitemShut
  {NoStop}%
\bibitem [{\citenamefont {S{\'a}}\ \emph {et~al.}(2020)\citenamefont {S{\'a}},
  \citenamefont {Ribeiro},\ and\ \citenamefont
  {Prosen}}]{Lucas2020spectralrandomLiouvillian}%
  \BibitemOpen
  \bibfield  {author} {\bibinfo {author} {\bibfnamefont {L.}~\bibnamefont
  {S{\'a}}}, \bibinfo {author} {\bibfnamefont {P.}~\bibnamefont {Ribeiro}}, \
  and\ \bibinfo {author} {\bibfnamefont {T.}~\bibnamefont {Prosen}},\
  }\bibfield  {title} {\enquote {\bibinfo {title} {Spectral and steady-state
  properties of random liouvillians},}\ }\href
  {https://iopscience.iop.org/article/10.1088/1751-8121/ab9337} {\bibfield
  {journal} {\bibinfo  {journal} {Journal of Physics A: Mathematical and
  Theoretical}\ } (\bibinfo {year} {2020})}\BibitemShut {NoStop}%
\bibitem [{\citenamefont {Ziolkowska}\ and\ \citenamefont
  {Essler}(2020)}]{ziolkowska2020yang}%
  \BibitemOpen
  \bibfield  {author} {\bibinfo {author} {\bibfnamefont {A.~A.}\ \bibnamefont
  {Ziolkowska}}\ and\ \bibinfo {author} {\bibfnamefont {F.~H.}\ \bibnamefont
  {Essler}},\ }\bibfield  {title} {\enquote {\bibinfo {title} {Yang-baxter
  integrable lindblad equations},}\ }\href
  {https://scipost.org/10.21468/SciPostPhys.8.3.044} {\bibfield  {journal}
  {\bibinfo  {journal} {SciPost Physics}\ }\textbf {\bibinfo {volume} {8}},\
  \bibinfo {pages} {044} (\bibinfo {year} {2020})}\BibitemShut {NoStop}%
\bibitem [{\citenamefont {Nakagawa}\ \emph {et~al.}(2020)\citenamefont
  {Nakagawa}, \citenamefont {Kawakami},\ and\ \citenamefont
  {Ueda}}]{nakagawa2020exact}%
  \BibitemOpen
  \bibfield  {author} {\bibinfo {author} {\bibfnamefont {M.}~\bibnamefont
  {Nakagawa}}, \bibinfo {author} {\bibfnamefont {N.}~\bibnamefont {Kawakami}},
  \ and\ \bibinfo {author} {\bibfnamefont {M.}~\bibnamefont {Ueda}},\
  }\bibfield  {title} {\enquote {\bibinfo {title} {Exact liouvillian spectrum
  of a one-dimensional dissipative hubbard model},}\ }\href
  {https://arxiv.org/abs/2003.14202} {\bibfield  {journal} {\bibinfo  {journal}
  {arXiv:2003.14202}\ } (\bibinfo {year} {2020})}\BibitemShut {NoStop}%
\bibitem [{\citenamefont {Bu{\v{c}}a}\ \emph {et~al.}(2020)\citenamefont
  {Bu{\v{c}}a}, \citenamefont {Booker}, \citenamefont {Medenjak},\ and\
  \citenamefont {Jaksch}}]{buvca2020bethe}%
  \BibitemOpen
  \bibfield  {author} {\bibinfo {author} {\bibfnamefont {B.}~\bibnamefont
  {Bu{\v{c}}a}}, \bibinfo {author} {\bibfnamefont {C.}~\bibnamefont {Booker}},
  \bibinfo {author} {\bibfnamefont {M.}~\bibnamefont {Medenjak}}, \ and\
  \bibinfo {author} {\bibfnamefont {D.}~\bibnamefont {Jaksch}},\ }\bibfield
  {title} {\enquote {\bibinfo {title} {Bethe ansatz approach for dissipation:
  exact solutions of quantum many-body dynamics under loss},}\ }\href
  {https://doi.org/10.1088/1367-2630/abd124} {\bibfield  {journal} {\bibinfo
  {journal} {New Journal of Physics}\ }\textbf {\bibinfo {volume} {22}},\
  \bibinfo {pages} {123040} (\bibinfo {year} {2020})}\BibitemShut {NoStop}%
\bibitem [{\citenamefont {Lloyd}(1993)}]{Lloyd1993Quantum}%
  \BibitemOpen
  \bibfield  {author} {\bibinfo {author} {\bibfnamefont {S.}~\bibnamefont
  {Lloyd}},\ }\bibfield  {title} {\enquote {\bibinfo {title}
  {Quantum-mechanical computers and uncomputability},}\ }\href {\doibase
  10.1103/PhysRevLett.71.943} {\bibfield  {journal} {\bibinfo  {journal} {Phys.
  Rev. Lett.}\ }\textbf {\bibinfo {volume} {71}},\ \bibinfo {pages} {943}
  (\bibinfo {year} {1993})}\BibitemShut {NoStop}%
\bibitem [{\citenamefont {Cubitt}\ \emph {et~al.}(2015)\citenamefont {Cubitt},
  \citenamefont {Perez-Garcia},\ and\ \citenamefont
  {Wolf}}]{Cubitt2015Undecidability}%
  \BibitemOpen
  \bibfield  {author} {\bibinfo {author} {\bibfnamefont {T.~S.}\ \bibnamefont
  {Cubitt}}, \bibinfo {author} {\bibfnamefont {D.}~\bibnamefont
  {Perez-Garcia}}, \ and\ \bibinfo {author} {\bibfnamefont {M.~M.}\
  \bibnamefont {Wolf}},\ }\bibfield  {title} {\enquote {\bibinfo {title}
  {Undecidability of the spectral gap},}\ }\href
  {https://doi.org/10.1038/nature16059} {\bibfield  {journal} {\bibinfo
  {journal} {Nature}\ }\textbf {\bibinfo {volume} {528}},\ \bibinfo {pages}
  {207} (\bibinfo {year} {2015})}\BibitemShut {NoStop}%
\bibitem [{\citenamefont {Bausch}\ \emph {et~al.}(2020)\citenamefont {Bausch},
  \citenamefont {Cubitt}, \citenamefont {Lucia},\ and\ \citenamefont
  {Perez-Garcia}}]{Bausch2020Undecidability}%
  \BibitemOpen
  \bibfield  {author} {\bibinfo {author} {\bibfnamefont {J.}~\bibnamefont
  {Bausch}}, \bibinfo {author} {\bibfnamefont {T.~S.}\ \bibnamefont {Cubitt}},
  \bibinfo {author} {\bibfnamefont {A.}~\bibnamefont {Lucia}}, \ and\ \bibinfo
  {author} {\bibfnamefont {D.}~\bibnamefont {Perez-Garcia}},\ }\bibfield
  {title} {\enquote {\bibinfo {title} {Undecidability of the spectral gap in
  one dimension},}\ }\href {\doibase 10.1103/PhysRevX.10.031038} {\bibfield
  {journal} {\bibinfo  {journal} {Phys. Rev. X}\ }\textbf {\bibinfo {volume}
  {10}},\ \bibinfo {pages} {031038} (\bibinfo {year} {2020})}\BibitemShut
  {NoStop}%
\bibitem [{\citenamefont {Weimer}\ \emph {et~al.}(2019)\citenamefont {Weimer},
  \citenamefont {Kshetrimayum},\ and\ \citenamefont
  {Or{\'u}s}}]{weimer2019simulationRMP}%
  \BibitemOpen
  \bibfield  {author} {\bibinfo {author} {\bibfnamefont {H.}~\bibnamefont
  {Weimer}}, \bibinfo {author} {\bibfnamefont {A.}~\bibnamefont
  {Kshetrimayum}}, \ and\ \bibinfo {author} {\bibfnamefont {R.}~\bibnamefont
  {Or{\'u}s}},\ }\bibfield  {title} {\enquote {\bibinfo {title} {Simulation
  methods for open quantum many-body systems},}\ }\href
  {https://arxiv.org/abs/1907.07079} {\bibfield  {journal} {\bibinfo  {journal}
  {arXiv:1907.07079}\ } (\bibinfo {year} {2019})}\BibitemShut {NoStop}%
\bibitem [{\citenamefont {Verstraete}\ \emph {et~al.}(2004)\citenamefont
  {Verstraete}, \citenamefont {Garc\'{\i}a-Ripoll},\ and\ \citenamefont
  {Cirac}}]{Verstraete2004Matrix}%
  \BibitemOpen
  \bibfield  {author} {\bibinfo {author} {\bibfnamefont {F.}~\bibnamefont
  {Verstraete}}, \bibinfo {author} {\bibfnamefont {J.~J.}\ \bibnamefont
  {Garc\'{\i}a-Ripoll}}, \ and\ \bibinfo {author} {\bibfnamefont {J.~I.}\
  \bibnamefont {Cirac}},\ }\bibfield  {title} {\enquote {\bibinfo {title}
  {Matrix product density operators: Simulation of finite-temperature and
  dissipative systems},}\ }\href {\doibase 10.1103/PhysRevLett.93.207204}
  {\bibfield  {journal} {\bibinfo  {journal} {Phys. Rev. Lett.}\ }\textbf
  {\bibinfo {volume} {93}},\ \bibinfo {pages} {207204} (\bibinfo {year}
  {2004})}\BibitemShut {NoStop}%
\bibitem [{\citenamefont {Zwolak}\ and\ \citenamefont
  {Vidal}(2004)}]{Zwolak2004Mixed}%
  \BibitemOpen
  \bibfield  {author} {\bibinfo {author} {\bibfnamefont {M.}~\bibnamefont
  {Zwolak}}\ and\ \bibinfo {author} {\bibfnamefont {G.}~\bibnamefont {Vidal}},\
  }\bibfield  {title} {\enquote {\bibinfo {title} {Mixed-state dynamics in
  one-dimensional quantum lattice systems: A time-dependent superoperator
  renormalization algorithm},}\ }\href {\doibase 10.1103/PhysRevLett.93.207205}
  {\bibfield  {journal} {\bibinfo  {journal} {Phys. Rev. Lett.}\ }\textbf
  {\bibinfo {volume} {93}},\ \bibinfo {pages} {207205} (\bibinfo {year}
  {2004})}\BibitemShut {NoStop}%
\bibitem [{\citenamefont {Cai}\ and\ \citenamefont
  {Barthel}(2013)}]{ZiCai2013AlgebraicversusExponential}%
  \BibitemOpen
  \bibfield  {author} {\bibinfo {author} {\bibfnamefont {Z.}~\bibnamefont
  {Cai}}\ and\ \bibinfo {author} {\bibfnamefont {T.}~\bibnamefont {Barthel}},\
  }\bibfield  {title} {\enquote {\bibinfo {title} {Algebraic versus exponential
  decoherence in dissipative many-particle systems},}\ }\href {\doibase
  10.1103/PhysRevLett.111.150403} {\bibfield  {journal} {\bibinfo  {journal}
  {Phys. Rev. Lett.}\ }\textbf {\bibinfo {volume} {111}},\ \bibinfo {pages}
  {150403} (\bibinfo {year} {2013})}\BibitemShut {NoStop}%
\bibitem [{\citenamefont {Mascarenhas}\ \emph {et~al.}(2015)\citenamefont
  {Mascarenhas}, \citenamefont {Flayac},\ and\ \citenamefont
  {Savona}}]{Mascarenhas2015Matrix}%
  \BibitemOpen
  \bibfield  {author} {\bibinfo {author} {\bibfnamefont {E.}~\bibnamefont
  {Mascarenhas}}, \bibinfo {author} {\bibfnamefont {H.}~\bibnamefont {Flayac}},
  \ and\ \bibinfo {author} {\bibfnamefont {V.}~\bibnamefont {Savona}},\
  }\bibfield  {title} {\enquote {\bibinfo {title} {Matrix-product-operator
  approach to the nonequilibrium steady state of driven-dissipative quantum
  arrays},}\ }\href {\doibase 10.1103/PhysRevA.92.022116} {\bibfield  {journal}
  {\bibinfo  {journal} {Phys. Rev. A}\ }\textbf {\bibinfo {volume} {92}},\
  \bibinfo {pages} {022116} (\bibinfo {year} {2015})}\BibitemShut {NoStop}%
\bibitem [{\citenamefont {Cui}\ \emph {et~al.}(2015)\citenamefont {Cui},
  \citenamefont {Cirac},\ and\ \citenamefont {Ba\~nuls}}]{Cui2015Variational}%
  \BibitemOpen
  \bibfield  {author} {\bibinfo {author} {\bibfnamefont {J.}~\bibnamefont
  {Cui}}, \bibinfo {author} {\bibfnamefont {J.~I.}\ \bibnamefont {Cirac}}, \
  and\ \bibinfo {author} {\bibfnamefont {M.~C.}\ \bibnamefont {Ba\~nuls}},\
  }\bibfield  {title} {\enquote {\bibinfo {title} {Variational matrix product
  operators for the steady state of dissipative quantum systems},}\ }\href
  {\doibase 10.1103/PhysRevLett.114.220601} {\bibfield  {journal} {\bibinfo
  {journal} {Phys. Rev. Lett.}\ }\textbf {\bibinfo {volume} {114}},\ \bibinfo
  {pages} {220601} (\bibinfo {year} {2015})}\BibitemShut {NoStop}%
\bibitem [{\citenamefont {\ifmmode \check{Z}\else
  \v{Z}\fi{}nidari\ifmmode~\check{c}\else
  \v{c}\fi{}}(2015)}]{Marko2015Lgapscaling}%
  \BibitemOpen
  \bibfield  {author} {\bibinfo {author} {\bibfnamefont {M.}~\bibnamefont
  {\ifmmode \check{Z}\else \v{Z}\fi{}nidari\ifmmode~\check{c}\else
  \v{c}\fi{}}},\ }\bibfield  {title} {\enquote {\bibinfo {title} {Relaxation
  times of dissipative many-body quantum systems},}\ }\href {\doibase
  10.1103/PhysRevE.92.042143} {\bibfield  {journal} {\bibinfo  {journal} {Phys.
  Rev. E}\ }\textbf {\bibinfo {volume} {92}},\ \bibinfo {pages} {042143}
  (\bibinfo {year} {2015})}\BibitemShut {NoStop}%
\bibitem [{\citenamefont {Werner}\ \emph {et~al.}(2016)\citenamefont {Werner},
  \citenamefont {Jaschke}, \citenamefont {Silvi}, \citenamefont {Kliesch},
  \citenamefont {Calarco}, \citenamefont {Eisert},\ and\ \citenamefont
  {Montangero}}]{Werner2016Positive}%
  \BibitemOpen
  \bibfield  {author} {\bibinfo {author} {\bibfnamefont {A.~H.}\ \bibnamefont
  {Werner}}, \bibinfo {author} {\bibfnamefont {D.}~\bibnamefont {Jaschke}},
  \bibinfo {author} {\bibfnamefont {P.}~\bibnamefont {Silvi}}, \bibinfo
  {author} {\bibfnamefont {M.}~\bibnamefont {Kliesch}}, \bibinfo {author}
  {\bibfnamefont {T.}~\bibnamefont {Calarco}}, \bibinfo {author} {\bibfnamefont
  {J.}~\bibnamefont {Eisert}}, \ and\ \bibinfo {author} {\bibfnamefont
  {S.}~\bibnamefont {Montangero}},\ }\bibfield  {title} {\enquote {\bibinfo
  {title} {Positive tensor network approach for simulating open quantum
  many-body systems},}\ }\href {\doibase 10.1103/PhysRevLett.116.237201}
  {\bibfield  {journal} {\bibinfo  {journal} {Phys. Rev. Lett.}\ }\textbf
  {\bibinfo {volume} {116}},\ \bibinfo {pages} {237201} (\bibinfo {year}
  {2016})}\BibitemShut {NoStop}%
\bibitem [{\citenamefont {Gangat}\ \emph {et~al.}(2017)\citenamefont {Gangat},
  \citenamefont {I},\ and\ \citenamefont {Kao}}]{Gangat2017Steady}%
  \BibitemOpen
  \bibfield  {author} {\bibinfo {author} {\bibfnamefont {A.~A.}\ \bibnamefont
  {Gangat}}, \bibinfo {author} {\bibfnamefont {T.}~\bibnamefont {I}}, \ and\
  \bibinfo {author} {\bibfnamefont {Y.-J.}\ \bibnamefont {Kao}},\ }\bibfield
  {title} {\enquote {\bibinfo {title} {Steady states of infinite-size
  dissipative quantum chains via imaginary time evolution},}\ }\href {\doibase
  10.1103/PhysRevLett.119.010501} {\bibfield  {journal} {\bibinfo  {journal}
  {Phys. Rev. Lett.}\ }\textbf {\bibinfo {volume} {119}},\ \bibinfo {pages}
  {010501} (\bibinfo {year} {2017})}\BibitemShut {NoStop}%
\bibitem [{\citenamefont {Kshetrimayum}\ \emph {et~al.}(2017)\citenamefont
  {Kshetrimayum}, \citenamefont {Weimer},\ and\ \citenamefont
  {Or{\'u}s}}]{kshetrimayum2017simpleTNsteady}%
  \BibitemOpen
  \bibfield  {author} {\bibinfo {author} {\bibfnamefont {A.}~\bibnamefont
  {Kshetrimayum}}, \bibinfo {author} {\bibfnamefont {H.}~\bibnamefont
  {Weimer}}, \ and\ \bibinfo {author} {\bibfnamefont {R.}~\bibnamefont
  {Or{\'u}s}},\ }\bibfield  {title} {\enquote {\bibinfo {title} {A simple
  tensor network algorithm for two-dimensional steady states},}\ }\href
  {https://doi.org/10.1038/s41467-017-01511-6} {\bibfield  {journal} {\bibinfo
  {journal} {Nat. Commun.}\ }\textbf {\bibinfo {volume} {8}},\ \bibinfo {pages}
  {1} (\bibinfo {year} {2017})}\BibitemShut {NoStop}%
\bibitem [{\citenamefont {Finazzi}\ \emph {et~al.}(2015)\citenamefont
  {Finazzi}, \citenamefont {Le~Boit\'e}, \citenamefont {Storme}, \citenamefont
  {Baksic},\ and\ \citenamefont {Ciuti}}]{FinazziCorner}%
  \BibitemOpen
  \bibfield  {author} {\bibinfo {author} {\bibfnamefont {S.}~\bibnamefont
  {Finazzi}}, \bibinfo {author} {\bibfnamefont {A.}~\bibnamefont {Le~Boit\'e}},
  \bibinfo {author} {\bibfnamefont {F.}~\bibnamefont {Storme}}, \bibinfo
  {author} {\bibfnamefont {A.}~\bibnamefont {Baksic}}, \ and\ \bibinfo {author}
  {\bibfnamefont {C.}~\bibnamefont {Ciuti}},\ }\bibfield  {title} {\enquote
  {\bibinfo {title} {Corner-space renormalization method for driven-dissipative
  two-dimensional correlated systems},}\ }\href {\doibase
  10.1103/PhysRevLett.115.080604} {\bibfield  {journal} {\bibinfo  {journal}
  {Phys. Rev. Lett.}\ }\textbf {\bibinfo {volume} {115}},\ \bibinfo {pages}
  {080604} (\bibinfo {year} {2015})}\BibitemShut {NoStop}%
\bibitem [{\citenamefont {Rota}\ \emph {et~al.}(2019)\citenamefont {Rota},
  \citenamefont {Minganti}, \citenamefont {Ciuti},\ and\ \citenamefont
  {Savona}}]{Rota2019Quantum}%
  \BibitemOpen
  \bibfield  {author} {\bibinfo {author} {\bibfnamefont {R.}~\bibnamefont
  {Rota}}, \bibinfo {author} {\bibfnamefont {F.}~\bibnamefont {Minganti}},
  \bibinfo {author} {\bibfnamefont {C.}~\bibnamefont {Ciuti}}, \ and\ \bibinfo
  {author} {\bibfnamefont {V.}~\bibnamefont {Savona}},\ }\bibfield  {title}
  {\enquote {\bibinfo {title} {Quantum critical regime in a quadratically
  driven nonlinear photonic lattice},}\ }\href {\doibase
  10.1103/PhysRevLett.122.110405} {\bibfield  {journal} {\bibinfo  {journal}
  {Phys. Rev. Lett.}\ }\textbf {\bibinfo {volume} {122}},\ \bibinfo {pages}
  {110405} (\bibinfo {year} {2019})}\BibitemShut {NoStop}%
\bibitem [{\citenamefont {Jin}\ \emph {et~al.}(2016)\citenamefont {Jin},
  \citenamefont {Biella}, \citenamefont {Viyuela}, \citenamefont {Mazza},
  \citenamefont {Keeling}, \citenamefont {Fazio},\ and\ \citenamefont
  {Rossini}}]{Jin2016Cluster}%
  \BibitemOpen
  \bibfield  {author} {\bibinfo {author} {\bibfnamefont {J.}~\bibnamefont
  {Jin}}, \bibinfo {author} {\bibfnamefont {A.}~\bibnamefont {Biella}},
  \bibinfo {author} {\bibfnamefont {O.}~\bibnamefont {Viyuela}}, \bibinfo
  {author} {\bibfnamefont {L.}~\bibnamefont {Mazza}}, \bibinfo {author}
  {\bibfnamefont {J.}~\bibnamefont {Keeling}}, \bibinfo {author} {\bibfnamefont
  {R.}~\bibnamefont {Fazio}}, \ and\ \bibinfo {author} {\bibfnamefont
  {D.}~\bibnamefont {Rossini}},\ }\bibfield  {title} {\enquote {\bibinfo
  {title} {Cluster mean-field approach to the steady-state phase diagram of
  dissipative spin systems},}\ }\href {\doibase 10.1103/PhysRevX.6.031011}
  {\bibfield  {journal} {\bibinfo  {journal} {Phys. Rev. X}\ }\textbf {\bibinfo
  {volume} {6}},\ \bibinfo {pages} {031011} (\bibinfo {year}
  {2016})}\BibitemShut {NoStop}%
\bibitem [{\citenamefont {Nagy}\ and\ \citenamefont
  {Savona}(2018)}]{Nagy2018Driven}%
  \BibitemOpen
  \bibfield  {author} {\bibinfo {author} {\bibfnamefont {A.}~\bibnamefont
  {Nagy}}\ and\ \bibinfo {author} {\bibfnamefont {V.}~\bibnamefont {Savona}},\
  }\bibfield  {title} {\enquote {\bibinfo {title} {Driven-dissipative quantum
  monte carlo method for open quantum systems},}\ }\href {\doibase
  10.1103/PhysRevA.97.052129} {\bibfield  {journal} {\bibinfo  {journal} {Phys.
  Rev. A}\ }\textbf {\bibinfo {volume} {97}},\ \bibinfo {pages} {052129}
  (\bibinfo {year} {2018})}\BibitemShut {NoStop}%
\bibitem [{\citenamefont {Yan}\ \emph {et~al.}(2018)\citenamefont {Yan},
  \citenamefont {Pollet}, \citenamefont {Lou}, \citenamefont {Wang},
  \citenamefont {Chen},\ and\ \citenamefont {Cai}}]{Yan2018Interacting}%
  \BibitemOpen
  \bibfield  {author} {\bibinfo {author} {\bibfnamefont {Z.}~\bibnamefont
  {Yan}}, \bibinfo {author} {\bibfnamefont {L.}~\bibnamefont {Pollet}},
  \bibinfo {author} {\bibfnamefont {J.}~\bibnamefont {Lou}}, \bibinfo {author}
  {\bibfnamefont {X.}~\bibnamefont {Wang}}, \bibinfo {author} {\bibfnamefont
  {Y.}~\bibnamefont {Chen}}, \ and\ \bibinfo {author} {\bibfnamefont
  {Z.}~\bibnamefont {Cai}},\ }\bibfield  {title} {\enquote {\bibinfo {title}
  {Interacting lattice systems with quantum dissipation: A quantum monte carlo
  study},}\ }\href {\doibase 10.1103/PhysRevB.97.035148} {\bibfield  {journal}
  {\bibinfo  {journal} {Phys. Rev. B}\ }\textbf {\bibinfo {volume} {97}},\
  \bibinfo {pages} {035148} (\bibinfo {year} {2018})}\BibitemShut {NoStop}%
\bibitem [{\citenamefont {Casteels}\ \emph {et~al.}(2018)\citenamefont
  {Casteels}, \citenamefont {Wilson},\ and\ \citenamefont
  {Wouters}}]{Casteels2018Gutzwiller}%
  \BibitemOpen
  \bibfield  {author} {\bibinfo {author} {\bibfnamefont {W.}~\bibnamefont
  {Casteels}}, \bibinfo {author} {\bibfnamefont {R.~M.}\ \bibnamefont
  {Wilson}}, \ and\ \bibinfo {author} {\bibfnamefont {M.}~\bibnamefont
  {Wouters}},\ }\bibfield  {title} {\enquote {\bibinfo {title} {Gutzwiller
  monte carlo approach for a critical dissipative spin model},}\ }\href
  {\doibase 10.1103/PhysRevA.97.062107} {\bibfield  {journal} {\bibinfo
  {journal} {Phys. Rev. A}\ }\textbf {\bibinfo {volume} {97}},\ \bibinfo
  {pages} {062107} (\bibinfo {year} {2018})}\BibitemShut {NoStop}%
\bibitem [{\citenamefont {Goodfellow}\ \emph {et~al.}(2016)\citenamefont
  {Goodfellow}, \citenamefont {Bengio},\ and\ \citenamefont
  {Courville}}]{goodfellow2016deep}%
  \BibitemOpen
  \bibfield  {author} {\bibinfo {author} {\bibfnamefont {I.}~\bibnamefont
  {Goodfellow}}, \bibinfo {author} {\bibfnamefont {Y.}~\bibnamefont {Bengio}},
  \ and\ \bibinfo {author} {\bibfnamefont {A.}~\bibnamefont {Courville}},\
  }\href@noop {} {\emph {\bibinfo {title} {Deep learning}}}\ (\bibinfo
  {publisher} {MIT press},\ \bibinfo {year} {2016})\BibitemShut {NoStop}%
\bibitem [{\citenamefont {Hartmann}\ and\ \citenamefont
  {Carleo}(2019)}]{Hartmann2019Neural}%
  \BibitemOpen
  \bibfield  {author} {\bibinfo {author} {\bibfnamefont {M.~J.}\ \bibnamefont
  {Hartmann}}\ and\ \bibinfo {author} {\bibfnamefont {G.}~\bibnamefont
  {Carleo}},\ }\bibfield  {title} {\enquote {\bibinfo {title} {Neural-network
  approach to dissipative quantum many-body dynamics},}\ }\href {\doibase
  10.1103/PhysRevLett.122.250502} {\bibfield  {journal} {\bibinfo  {journal}
  {Phys. Rev. Lett.}\ }\textbf {\bibinfo {volume} {122}},\ \bibinfo {pages}
  {250502} (\bibinfo {year} {2019})}\BibitemShut {NoStop}%
\bibitem [{\citenamefont {Vicentini}\ \emph {et~al.}(2019)\citenamefont
  {Vicentini}, \citenamefont {Biella}, \citenamefont {Regnault},\ and\
  \citenamefont {Ciuti}}]{Vicentini2019Variational}%
  \BibitemOpen
  \bibfield  {author} {\bibinfo {author} {\bibfnamefont {F.}~\bibnamefont
  {Vicentini}}, \bibinfo {author} {\bibfnamefont {A.}~\bibnamefont {Biella}},
  \bibinfo {author} {\bibfnamefont {N.}~\bibnamefont {Regnault}}, \ and\
  \bibinfo {author} {\bibfnamefont {C.}~\bibnamefont {Ciuti}},\ }\bibfield
  {title} {\enquote {\bibinfo {title} {Variational neural-network ansatz for
  steady states in open quantum systems},}\ }\href {\doibase
  10.1103/PhysRevLett.122.250503} {\bibfield  {journal} {\bibinfo  {journal}
  {Phys. Rev. Lett.}\ }\textbf {\bibinfo {volume} {122}},\ \bibinfo {pages}
  {250503} (\bibinfo {year} {2019})}\BibitemShut {NoStop}%
\bibitem [{\citenamefont {Nagy}\ and\ \citenamefont
  {Savona}(2019)}]{Nagy2019Variational}%
  \BibitemOpen
  \bibfield  {author} {\bibinfo {author} {\bibfnamefont {A.}~\bibnamefont
  {Nagy}}\ and\ \bibinfo {author} {\bibfnamefont {V.}~\bibnamefont {Savona}},\
  }\bibfield  {title} {\enquote {\bibinfo {title} {Variational quantum monte
  carlo method with a neural-network ansatz for open quantum systems},}\ }\href
  {\doibase 10.1103/PhysRevLett.122.250501} {\bibfield  {journal} {\bibinfo
  {journal} {Phys. Rev. Lett.}\ }\textbf {\bibinfo {volume} {122}},\ \bibinfo
  {pages} {250501} (\bibinfo {year} {2019})}\BibitemShut {NoStop}%
\bibitem [{\citenamefont {Yoshioka}\ and\ \citenamefont
  {Hamazaki}(2019)}]{Yoshioka2019Constructing}%
  \BibitemOpen
  \bibfield  {author} {\bibinfo {author} {\bibfnamefont {N.}~\bibnamefont
  {Yoshioka}}\ and\ \bibinfo {author} {\bibfnamefont {R.}~\bibnamefont
  {Hamazaki}},\ }\bibfield  {title} {\enquote {\bibinfo {title} {Constructing
  neural stationary states for open quantum many-body systems},}\ }\href
  {\doibase 10.1103/PhysRevB.99.214306} {\bibfield  {journal} {\bibinfo
  {journal} {Phys. Rev. B}\ }\textbf {\bibinfo {volume} {99}},\ \bibinfo
  {pages} {214306} (\bibinfo {year} {2019})}\BibitemShut {NoStop}%
\bibitem [{\citenamefont {Carleo}\ and\ \citenamefont
  {Troyer}(2017)}]{Carleo2016Solving}%
  \BibitemOpen
  \bibfield  {author} {\bibinfo {author} {\bibfnamefont {G.}~\bibnamefont
  {Carleo}}\ and\ \bibinfo {author} {\bibfnamefont {M.}~\bibnamefont
  {Troyer}},\ }\bibfield  {title} {\enquote {\bibinfo {title} {Solving the
  quantum many-body problem with artificial neural networks},}\ }\href
  {\doibase 10.1126/science.aag2302} {\bibfield  {journal} {\bibinfo  {journal}
  {Science}\ }\textbf {\bibinfo {volume} {355}},\ \bibinfo {pages} {602}
  (\bibinfo {year} {2017})}\BibitemShut {NoStop}%
\bibitem [{\citenamefont {Torlai}\ and\ \citenamefont
  {Melko}(2018)}]{Torlai2018Latent}%
  \BibitemOpen
  \bibfield  {author} {\bibinfo {author} {\bibfnamefont {G.}~\bibnamefont
  {Torlai}}\ and\ \bibinfo {author} {\bibfnamefont {R.~G.}\ \bibnamefont
  {Melko}},\ }\bibfield  {title} {\enquote {\bibinfo {title} {Latent space
  purification via neural density operators},}\ }\href {\doibase
  10.1103/PhysRevLett.120.240503} {\bibfield  {journal} {\bibinfo  {journal}
  {Phys. Rev. Lett.}\ }\textbf {\bibinfo {volume} {120}},\ \bibinfo {pages}
  {240503} (\bibinfo {year} {2018})}\BibitemShut {NoStop}%
\bibitem [{\citenamefont {Carrasquilla}\ \emph {et~al.}(2019)\citenamefont
  {Carrasquilla}, \citenamefont {Torlai}, \citenamefont {Melko},\ and\
  \citenamefont {Aolita}}]{Carrasquilla2019Reconstructing}%
  \BibitemOpen
  \bibfield  {author} {\bibinfo {author} {\bibfnamefont {J.}~\bibnamefont
  {Carrasquilla}}, \bibinfo {author} {\bibfnamefont {G.}~\bibnamefont
  {Torlai}}, \bibinfo {author} {\bibfnamefont {R.~G.}\ \bibnamefont {Melko}}, \
  and\ \bibinfo {author} {\bibfnamefont {L.}~\bibnamefont {Aolita}},\
  }\bibfield  {title} {\enquote {\bibinfo {title} {Reconstructing quantum
  states with generative models},}\ }\href
  {https://www.nature.com/articles/s42256-019-0028-1} {\bibfield  {journal}
  {\bibinfo  {journal} {Nature Machine Intelligence}\ }\textbf {\bibinfo
  {volume} {1}},\ \bibinfo {pages} {155} (\bibinfo {year} {2019})}\BibitemShut
  {NoStop}%
\bibitem [{\citenamefont {Banchi}\ \emph {et~al.}(2018)\citenamefont {Banchi},
  \citenamefont {Grant}, \citenamefont {Rocchetto},\ and\ \citenamefont
  {Severini}}]{Banchi2018Modelling}%
  \BibitemOpen
  \bibfield  {author} {\bibinfo {author} {\bibfnamefont {L.}~\bibnamefont
  {Banchi}}, \bibinfo {author} {\bibfnamefont {E.}~\bibnamefont {Grant}},
  \bibinfo {author} {\bibfnamefont {A.}~\bibnamefont {Rocchetto}}, \ and\
  \bibinfo {author} {\bibfnamefont {S.}~\bibnamefont {Severini}},\ }\bibfield
  {title} {\enquote {\bibinfo {title} {Modelling non-markovian quantum
  processes with recurrent neural networks},}\ }\href
  {https://iopscience.iop.org/article/10.1088/1367-2630/aaf749/meta} {\bibfield
   {journal} {\bibinfo  {journal} {New Journal of Physics}\ }\textbf {\bibinfo
  {volume} {20}},\ \bibinfo {pages} {123030} (\bibinfo {year}
  {2018})}\BibitemShut {NoStop}%
\bibitem [{\citenamefont {Cai}\ and\ \citenamefont
  {Liu}(2018)}]{Cai2018Approximating}%
  \BibitemOpen
  \bibfield  {author} {\bibinfo {author} {\bibfnamefont {Z.}~\bibnamefont
  {Cai}}\ and\ \bibinfo {author} {\bibfnamefont {J.}~\bibnamefont {Liu}},\
  }\bibfield  {title} {\enquote {\bibinfo {title} {Approximating quantum
  many-body wave functions using artificial neural networks},}\ }\href
  {\doibase 10.1103/PhysRevB.97.035116} {\bibfield  {journal} {\bibinfo
  {journal} {Phys. Rev. B}\ }\textbf {\bibinfo {volume} {97}},\ \bibinfo
  {pages} {035116} (\bibinfo {year} {2018})}\BibitemShut {NoStop}%
\bibitem [{\citenamefont {Schmitt}\ and\ \citenamefont
  {Heyl}(2020)}]{Markus2020manybodydynamics2D}%
  \BibitemOpen
  \bibfield  {author} {\bibinfo {author} {\bibfnamefont {M.}~\bibnamefont
  {Schmitt}}\ and\ \bibinfo {author} {\bibfnamefont {M.}~\bibnamefont {Heyl}},\
  }\bibfield  {title} {\enquote {\bibinfo {title} {Quantum many-body dynamics
  in two dimensions with artificial neural networks},}\ }\href {\doibase
  10.1103/PhysRevLett.125.100503} {\bibfield  {journal} {\bibinfo  {journal}
  {Phys. Rev. Lett.}\ }\textbf {\bibinfo {volume} {125}},\ \bibinfo {pages}
  {100503} (\bibinfo {year} {2020})}\BibitemShut {NoStop}%
\bibitem [{\citenamefont {Sorella}\ \emph {et~al.}(2007)\citenamefont
  {Sorella}, \citenamefont {Casula},\ and\ \citenamefont
  {Rocca}}]{Sorella2007Weak}%
  \BibitemOpen
  \bibfield  {author} {\bibinfo {author} {\bibfnamefont {S.}~\bibnamefont
  {Sorella}}, \bibinfo {author} {\bibfnamefont {M.}~\bibnamefont {Casula}}, \
  and\ \bibinfo {author} {\bibfnamefont {D.}~\bibnamefont {Rocca}},\ }\bibfield
   {title} {\enquote {\bibinfo {title} {Weak binding between two aromatic
  rings: Feeling the van der waals attraction by quantum monte carlo
  methods},}\ }\href {\doibase 10.1063/1.2746035} {\bibfield  {journal}
  {\bibinfo  {journal} {J. Chem. Phys.}\ }\textbf {\bibinfo {volume} {127}},\
  \bibinfo {pages} {014105} (\bibinfo {year} {2007})}\BibitemShut {NoStop}%
\bibitem [{\citenamefont {Deng}\ \emph {et~al.}(2017)\citenamefont {Deng},
  \citenamefont {Li},\ and\ \citenamefont {Das~Sarma}}]{Deng2017Quantum}%
  \BibitemOpen
  \bibfield  {author} {\bibinfo {author} {\bibfnamefont {D.-L.}\ \bibnamefont
  {Deng}}, \bibinfo {author} {\bibfnamefont {X.}~\bibnamefont {Li}}, \ and\
  \bibinfo {author} {\bibfnamefont {S.}~\bibnamefont {Das~Sarma}},\ }\bibfield
  {title} {\enquote {\bibinfo {title} {Quantum entanglement in neural network
  states},}\ }\href {\doibase 10.1103/PhysRevX.7.021021} {\bibfield  {journal}
  {\bibinfo  {journal} {Phys. Rev. X}\ }\textbf {\bibinfo {volume} {7}},\
  \bibinfo {pages} {021021} (\bibinfo {year} {2017})}\BibitemShut {NoStop}%
\bibitem [{\citenamefont {Rossatto}\ and\ \citenamefont
  {Villas-Boas}(2016)}]{Rossatto2016Relaxationtime}%
  \BibitemOpen
  \bibfield  {author} {\bibinfo {author} {\bibfnamefont {D.~Z.}\ \bibnamefont
  {Rossatto}}\ and\ \bibinfo {author} {\bibfnamefont {C.~J.}\ \bibnamefont
  {Villas-Boas}},\ }\bibfield  {title} {\enquote {\bibinfo {title} {Relaxation
  time for monitoring the quantumness of an intense cavity field},}\ }\href
  {\doibase 10.1103/PhysRevA.94.033819} {\bibfield  {journal} {\bibinfo
  {journal} {Phys. Rev. A}\ }\textbf {\bibinfo {volume} {94}},\ \bibinfo
  {pages} {033819} (\bibinfo {year} {2016})}\BibitemShut {NoStop}%
\bibitem [{\citenamefont {Kessler}\ \emph {et~al.}(2012)\citenamefont
  {Kessler}, \citenamefont {Giedke}, \citenamefont {Imamoglu}, \citenamefont
  {Yelin}, \citenamefont {Lukin},\ and\ \citenamefont
  {Cirac}}]{Kessler2012Dissipative}%
  \BibitemOpen
  \bibfield  {author} {\bibinfo {author} {\bibfnamefont {E.~M.}\ \bibnamefont
  {Kessler}}, \bibinfo {author} {\bibfnamefont {G.}~\bibnamefont {Giedke}},
  \bibinfo {author} {\bibfnamefont {A.}~\bibnamefont {Imamoglu}}, \bibinfo
  {author} {\bibfnamefont {S.~F.}\ \bibnamefont {Yelin}}, \bibinfo {author}
  {\bibfnamefont {M.~D.}\ \bibnamefont {Lukin}}, \ and\ \bibinfo {author}
  {\bibfnamefont {J.~I.}\ \bibnamefont {Cirac}},\ }\bibfield  {title} {\enquote
  {\bibinfo {title} {Dissipative phase transition in a central spin system},}\
  }\href {\doibase 10.1103/PhysRevA.86.012116} {\bibfield  {journal} {\bibinfo
  {journal} {Phys. Rev. A}\ }\textbf {\bibinfo {volume} {86}},\ \bibinfo
  {pages} {012116} (\bibinfo {year} {2012})}\BibitemShut {NoStop}%
\bibitem [{\citenamefont {Minganti}\ \emph {et~al.}(2018)\citenamefont
  {Minganti}, \citenamefont {Biella}, \citenamefont {Bartolo},\ and\
  \citenamefont {Ciuti}}]{Minganti2018Spectral}%
  \BibitemOpen
  \bibfield  {author} {\bibinfo {author} {\bibfnamefont {F.}~\bibnamefont
  {Minganti}}, \bibinfo {author} {\bibfnamefont {A.}~\bibnamefont {Biella}},
  \bibinfo {author} {\bibfnamefont {N.}~\bibnamefont {Bartolo}}, \ and\
  \bibinfo {author} {\bibfnamefont {C.}~\bibnamefont {Ciuti}},\ }\bibfield
  {title} {\enquote {\bibinfo {title} {Spectral theory of liouvillians for
  dissipative phase transitions},}\ }\href {\doibase
  10.1103/PhysRevA.98.042118} {\bibfield  {journal} {\bibinfo  {journal} {Phys.
  Rev. A}\ }\textbf {\bibinfo {volume} {98}},\ \bibinfo {pages} {042118}
  (\bibinfo {year} {2018})}\BibitemShut {NoStop}%
\bibitem [{\citenamefont {Song}\ \emph {et~al.}(2019)\citenamefont {Song},
  \citenamefont {Yao},\ and\ \citenamefont {Wang}}]{Song2019NonHermitian}%
  \BibitemOpen
  \bibfield  {author} {\bibinfo {author} {\bibfnamefont {F.}~\bibnamefont
  {Song}}, \bibinfo {author} {\bibfnamefont {S.}~\bibnamefont {Yao}}, \ and\
  \bibinfo {author} {\bibfnamefont {Z.}~\bibnamefont {Wang}},\ }\bibfield
  {title} {\enquote {\bibinfo {title} {Non-hermitian skin effect and chiral
  damping in open quantum systems},}\ }\href {\doibase
  10.1103/PhysRevLett.123.170401} {\bibfield  {journal} {\bibinfo  {journal}
  {Phys. Rev. Lett.}\ }\textbf {\bibinfo {volume} {123}},\ \bibinfo {pages}
  {170401} (\bibinfo {year} {2019})}\BibitemShut {NoStop}%
\bibitem [{LGv()}]{LGviaRBMSuppM}%
  \BibitemOpen
  \href@noop {} {}\bibinfo {note} {See Supplemental Material at [URL will be
  inserted by publisher] for details on the stochastic reconfiguration,
  computational cost, analytical solution of the full Liouvillian spectrum for
  the dissipative XXZ model, and more numerical data, which includes Refs.
  \cite{metropolis1953equation,melko2019restrictedreview,offdiagonalBetheAnsatz,Lukin2013MeanfieldclosedXYZ,Huybrechts2020meanfieldXYZ}}\BibitemShut
  {NoStop}%
\bibitem [{\citenamefont {Vieijra}\ \emph {et~al.}(2020)\citenamefont
  {Vieijra}, \citenamefont {Casert}, \citenamefont {Nys}, \citenamefont
  {De~Neve}, \citenamefont {Haegeman}, \citenamefont {Ryckebusch},\ and\
  \citenamefont {Verstraete}}]{Vieijra2020Restricted}%
  \BibitemOpen
  \bibfield  {author} {\bibinfo {author} {\bibfnamefont {T.}~\bibnamefont
  {Vieijra}}, \bibinfo {author} {\bibfnamefont {C.}~\bibnamefont {Casert}},
  \bibinfo {author} {\bibfnamefont {J.}~\bibnamefont {Nys}}, \bibinfo {author}
  {\bibfnamefont {W.}~\bibnamefont {De~Neve}}, \bibinfo {author} {\bibfnamefont
  {J.}~\bibnamefont {Haegeman}}, \bibinfo {author} {\bibfnamefont
  {J.}~\bibnamefont {Ryckebusch}}, \ and\ \bibinfo {author} {\bibfnamefont
  {F.}~\bibnamefont {Verstraete}},\ }\bibfield  {title} {\enquote {\bibinfo
  {title} {Restricted boltzmann machines for quantum states with non-abelian or
  anyonic symmetries},}\ }\href {\doibase 10.1103/PhysRevLett.124.097201}
  {\bibfield  {journal} {\bibinfo  {journal} {Phys. Rev. Lett.}\ }\textbf
  {\bibinfo {volume} {124}},\ \bibinfo {pages} {097201} (\bibinfo {year}
  {2020})}\BibitemShut {NoStop}%
\bibitem [{\citenamefont {Choo}\ \emph {et~al.}(2018)\citenamefont {Choo},
  \citenamefont {Carleo}, \citenamefont {Regnault},\ and\ \citenamefont
  {Neupert}}]{Choo2018Symmetries}%
  \BibitemOpen
  \bibfield  {author} {\bibinfo {author} {\bibfnamefont {K.}~\bibnamefont
  {Choo}}, \bibinfo {author} {\bibfnamefont {G.}~\bibnamefont {Carleo}},
  \bibinfo {author} {\bibfnamefont {N.}~\bibnamefont {Regnault}}, \ and\
  \bibinfo {author} {\bibfnamefont {T.}~\bibnamefont {Neupert}},\ }\bibfield
  {title} {\enquote {\bibinfo {title} {Symmetries and many-body excitations
  with neural-network quantum states},}\ }\href {\doibase
  10.1103/PhysRevLett.121.167204} {\bibfield  {journal} {\bibinfo  {journal}
  {Phys. Rev. Lett.}\ }\textbf {\bibinfo {volume} {121}},\ \bibinfo {pages}
  {167204} (\bibinfo {year} {2018})}\BibitemShut {NoStop}%
\bibitem [{\citenamefont {Lee}\ \emph {et~al.}(2013{\natexlab{a}})\citenamefont
  {Lee}, \citenamefont {Gopalakrishnan},\ and\ \citenamefont
  {Lukin}}]{Lee2013Unconventional}%
  \BibitemOpen
  \bibfield  {author} {\bibinfo {author} {\bibfnamefont {T.~E.}\ \bibnamefont
  {Lee}}, \bibinfo {author} {\bibfnamefont {S.}~\bibnamefont {Gopalakrishnan}},
  \ and\ \bibinfo {author} {\bibfnamefont {M.~D.}\ \bibnamefont {Lukin}},\
  }\bibfield  {title} {\enquote {\bibinfo {title} {Unconventional magnetism via
  optical pumping of interacting spin systems},}\ }\href {\doibase
  10.1103/PhysRevLett.110.257204} {\bibfield  {journal} {\bibinfo  {journal}
  {Phys. Rev. Lett.}\ }\textbf {\bibinfo {volume} {110}},\ \bibinfo {pages}
  {257204} (\bibinfo {year} {2013}{\natexlab{a}})}\BibitemShut {NoStop}%
\bibitem [{\citenamefont {Rota}\ \emph {et~al.}(2017)\citenamefont {Rota},
  \citenamefont {Storme}, \citenamefont {Bartolo}, \citenamefont {Fazio},\ and\
  \citenamefont {Ciuti}}]{Rota2017Critical}%
  \BibitemOpen
  \bibfield  {author} {\bibinfo {author} {\bibfnamefont {R.}~\bibnamefont
  {Rota}}, \bibinfo {author} {\bibfnamefont {F.}~\bibnamefont {Storme}},
  \bibinfo {author} {\bibfnamefont {N.}~\bibnamefont {Bartolo}}, \bibinfo
  {author} {\bibfnamefont {R.}~\bibnamefont {Fazio}}, \ and\ \bibinfo {author}
  {\bibfnamefont {C.}~\bibnamefont {Ciuti}},\ }\bibfield  {title} {\enquote
  {\bibinfo {title} {Critical behavior of dissipative two-dimensional spin
  lattices},}\ }\href {\doibase 10.1103/PhysRevB.95.134431} {\bibfield
  {journal} {\bibinfo  {journal} {Phys. Rev. B}\ }\textbf {\bibinfo {volume}
  {95}},\ \bibinfo {pages} {134431} (\bibinfo {year} {2017})}\BibitemShut
  {NoStop}%
\bibitem [{\citenamefont {Rota}\ \emph {et~al.}(2018)\citenamefont {Rota},
  \citenamefont {Minganti}, \citenamefont {Biella},\ and\ \citenamefont
  {Ciuti}}]{Rota2018Dynamical}%
  \BibitemOpen
  \bibfield  {author} {\bibinfo {author} {\bibfnamefont {R.}~\bibnamefont
  {Rota}}, \bibinfo {author} {\bibfnamefont {F.}~\bibnamefont {Minganti}},
  \bibinfo {author} {\bibfnamefont {A.}~\bibnamefont {Biella}}, \ and\ \bibinfo
  {author} {\bibfnamefont {C.}~\bibnamefont {Ciuti}},\ }\bibfield  {title}
  {\enquote {\bibinfo {title} {Dynamical properties of dissipative xyz
  heisenberg lattices},}\ }\href
  {https://iopscience.iop.org/article/10.1088/1367-2630/aab703/meta} {\bibfield
   {journal} {\bibinfo  {journal} {New Journal of Physics}\ }\textbf {\bibinfo
  {volume} {20}},\ \bibinfo {pages} {045003} (\bibinfo {year}
  {2018})}\BibitemShut {NoStop}%
\bibitem [{\citenamefont {Huybrechts}\ and\ \citenamefont
  {Wouters}(2019)}]{Huybrechts2019Cluster}%
  \BibitemOpen
  \bibfield  {author} {\bibinfo {author} {\bibfnamefont {D.}~\bibnamefont
  {Huybrechts}}\ and\ \bibinfo {author} {\bibfnamefont {M.}~\bibnamefont
  {Wouters}},\ }\bibfield  {title} {\enquote {\bibinfo {title} {Cluster methods
  for the description of a driven-dissipative spin model},}\ }\href {\doibase
  10.1103/PhysRevA.99.043841} {\bibfield  {journal} {\bibinfo  {journal} {Phys.
  Rev. A}\ }\textbf {\bibinfo {volume} {99}},\ \bibinfo {pages} {043841}
  (\bibinfo {year} {2019})}\BibitemShut {NoStop}%
\bibitem [{\citenamefont {Torres}(2014)}]{torre2014closedform}%
  \BibitemOpen
  \bibfield  {author} {\bibinfo {author} {\bibfnamefont {J.~M.}\ \bibnamefont
  {Torres}},\ }\bibfield  {title} {\enquote {\bibinfo {title} {Closed-form
  solution of lindblad master equations without gain},}\ }\href {\doibase
  10.1103/PhysRevA.89.052133} {\bibfield  {journal} {\bibinfo  {journal} {Phys.
  Rev. A}\ }\textbf {\bibinfo {volume} {89}},\ \bibinfo {pages} {052133}
  (\bibinfo {year} {2014})}\BibitemShut {NoStop}%
\bibitem [{\citenamefont {Saito}(2017)}]{Saito2017Solving}%
  \BibitemOpen
  \bibfield  {author} {\bibinfo {author} {\bibfnamefont {H.}~\bibnamefont
  {Saito}},\ }\bibfield  {title} {\enquote {\bibinfo {title} {Solving the
  bose--hubbard model with machine learning},}\ }\href {\doibase
  10.7566/JPSJ.86.093001} {\bibfield  {journal} {\bibinfo  {journal} {J. Phys.
  Soc. Jpn.}\ }\textbf {\bibinfo {volume} {86}},\ \bibinfo {pages} {093001}
  (\bibinfo {year} {2017})}\BibitemShut {NoStop}%
\bibitem [{\citenamefont {Nomura}\ \emph {et~al.}(2017)\citenamefont {Nomura},
  \citenamefont {Darmawan}, \citenamefont {Yamaji},\ and\ \citenamefont
  {Imada}}]{Nomura2017Restricted}%
  \BibitemOpen
  \bibfield  {author} {\bibinfo {author} {\bibfnamefont {Y.}~\bibnamefont
  {Nomura}}, \bibinfo {author} {\bibfnamefont {A.~S.}\ \bibnamefont
  {Darmawan}}, \bibinfo {author} {\bibfnamefont {Y.}~\bibnamefont {Yamaji}}, \
  and\ \bibinfo {author} {\bibfnamefont {M.}~\bibnamefont {Imada}},\ }\bibfield
   {title} {\enquote {\bibinfo {title} {Restricted boltzmann machine learning
  for solving strongly correlated quantum systems},}\ }\href {\doibase
  10.1103/PhysRevB.96.205152} {\bibfield  {journal} {\bibinfo  {journal} {Phys.
  Rev. B}\ }\textbf {\bibinfo {volume} {96}},\ \bibinfo {pages} {205152}
  (\bibinfo {year} {2017})}\BibitemShut {NoStop}%
\bibitem [{\citenamefont {Choo}\ \emph {et~al.}(2020)\citenamefont {Choo},
  \citenamefont {Mezzacapo},\ and\ \citenamefont {Carleo}}]{Choo2020Fermionic}%
  \BibitemOpen
  \bibfield  {author} {\bibinfo {author} {\bibfnamefont {K.}~\bibnamefont
  {Choo}}, \bibinfo {author} {\bibfnamefont {A.}~\bibnamefont {Mezzacapo}}, \
  and\ \bibinfo {author} {\bibfnamefont {G.}~\bibnamefont {Carleo}},\
  }\bibfield  {title} {\enquote {\bibinfo {title} {Fermionic neural-network
  states for ab-initio electronic structure},}\ }\href
  {https://www.nature.com/articles/s41467-020-15724-9} {\bibfield  {journal}
  {\bibinfo  {journal} {Nat. Commun.}\ }\textbf {\bibinfo {volume} {11}},\
  \bibinfo {pages} {1} (\bibinfo {year} {2020})}\BibitemShut {NoStop}%
\bibitem [{\citenamefont {Arora}\ and\ \citenamefont
  {Barak}(2009)}]{Arora2009Computational}%
  \BibitemOpen
  \bibfield  {author} {\bibinfo {author} {\bibfnamefont {S.}~\bibnamefont
  {Arora}}\ and\ \bibinfo {author} {\bibfnamefont {B.}~\bibnamefont {Barak}},\
  }\href@noop {} {\emph {\bibinfo {title} {Computational complexity: a modern
  approach}}}\ (\bibinfo  {publisher} {Cambridge University Press},\ \bibinfo
  {year} {2009})\BibitemShut {NoStop}%
\bibitem [{\citenamefont {Schollw\"ock}(2005)}]{Schollwock2005TheDMRG}%
  \BibitemOpen
  \bibfield  {author} {\bibinfo {author} {\bibfnamefont {U.}~\bibnamefont
  {Schollw\"ock}},\ }\bibfield  {title} {\enquote {\bibinfo {title} {The
  density-matrix renormalization group},}\ }\href {\doibase
  10.1103/RevModPhys.77.259} {\bibfield  {journal} {\bibinfo  {journal} {Rev.
  Mod. Phys.}\ }\textbf {\bibinfo {volume} {77}},\ \bibinfo {pages} {259}
  (\bibinfo {year} {2005})}\BibitemShut {NoStop}%
\bibitem [{\citenamefont {Gao}\ and\ \citenamefont
  {Duan}(2017)}]{Gao2017Efficient}%
  \BibitemOpen
  \bibfield  {author} {\bibinfo {author} {\bibfnamefont {X.}~\bibnamefont
  {Gao}}\ and\ \bibinfo {author} {\bibfnamefont {L.-M.}\ \bibnamefont {Duan}},\
  }\bibfield  {title} {\enquote {\bibinfo {title} {Efficient representation of
  quantum many-body states with deep neural networks},}\ }\href
  {https://www.nature.com/articles/s41467-017-00705-2} {\bibfield  {journal}
  {\bibinfo  {journal} {Nat. Commun.}\ }\textbf {\bibinfo {volume} {8}},\
  \bibinfo {pages} {1} (\bibinfo {year} {2017})}\BibitemShut {NoStop}%
\bibitem [{\citenamefont {Liu}\ \emph {et~al.}(2020)\citenamefont {Liu},
  \citenamefont {Duan},\ and\ \citenamefont {Deng}}]{Liu2020Solving}%
  \BibitemOpen
  \bibfield  {author} {\bibinfo {author} {\bibfnamefont {Z.}~\bibnamefont
  {Liu}}, \bibinfo {author} {\bibfnamefont {L.~M.}\ \bibnamefont {Duan}}, \
  and\ \bibinfo {author} {\bibfnamefont {D.-L.}\ \bibnamefont {Deng}},\
  }\bibfield  {title} {\enquote {\bibinfo {title} {Solving quantum master
  equations with deep quantum neural networks},}\ }\href
  {https://arxiv.org/abs/2008.05488} {\bibfield  {journal} {\bibinfo  {journal}
  {arXiv:2008.05488}\ } (\bibinfo {year} {2020})}\BibitemShut {NoStop}%
\bibitem [{\citenamefont {Preskill}(2018)}]{Preskill2018quantum}%
  \BibitemOpen
  \bibfield  {author} {\bibinfo {author} {\bibfnamefont {J.}~\bibnamefont
  {Preskill}},\ }\bibfield  {title} {\enquote {\bibinfo {title} {Quantum
  computing in the {NISQ} era and beyond},}\ }\href
  {https://doi.org/10.22331/q-2018-08-06-79} {\bibfield  {journal} {\bibinfo
  {journal} {{Quantum}}\ }\textbf {\bibinfo {volume} {2}},\ \bibinfo {pages}
  {79} (\bibinfo {year} {2018})}\BibitemShut {NoStop}%
\bibitem [{\citenamefont {Metropolis}\ \emph {et~al.}(1953)\citenamefont
  {Metropolis}, \citenamefont {Rosenbluth}, \citenamefont {Rosenbluth},
  \citenamefont {Teller},\ and\ \citenamefont
  {Teller}}]{metropolis1953equation}%
  \BibitemOpen
  \bibfield  {author} {\bibinfo {author} {\bibfnamefont {N.}~\bibnamefont
  {Metropolis}}, \bibinfo {author} {\bibfnamefont {A.~W.}\ \bibnamefont
  {Rosenbluth}}, \bibinfo {author} {\bibfnamefont {M.~N.}\ \bibnamefont
  {Rosenbluth}}, \bibinfo {author} {\bibfnamefont {A.~H.}\ \bibnamefont
  {Teller}}, \ and\ \bibinfo {author} {\bibfnamefont {E.}~\bibnamefont
  {Teller}},\ }\bibfield  {title} {\enquote {\bibinfo {title} {Equation of
  state calculations by fast computing machines},}\ }\href
  {https://aip.scitation.org/doi/abs/10.1063/1.1699114} {\bibfield  {journal}
  {\bibinfo  {journal} {J. Chem. Phys.}\ }\textbf {\bibinfo {volume} {21}},\
  \bibinfo {pages} {1087} (\bibinfo {year} {1953})}\BibitemShut {NoStop}%
\bibitem [{\citenamefont {Melko}\ \emph {et~al.}(2019)\citenamefont {Melko},
  \citenamefont {Carleo}, \citenamefont {Carrasquilla},\ and\ \citenamefont
  {Cirac}}]{melko2019restrictedreview}%
  \BibitemOpen
  \bibfield  {author} {\bibinfo {author} {\bibfnamefont {R.~G.}\ \bibnamefont
  {Melko}}, \bibinfo {author} {\bibfnamefont {G.}~\bibnamefont {Carleo}},
  \bibinfo {author} {\bibfnamefont {J.}~\bibnamefont {Carrasquilla}}, \ and\
  \bibinfo {author} {\bibfnamefont {J.~I.}\ \bibnamefont {Cirac}},\ }\bibfield
  {title} {\enquote {\bibinfo {title} {Restricted boltzmann machines in quantum
  physics},}\ }\href {https://www.nature.com/articles/s41567-019-0545-1}
  {\bibfield  {journal} {\bibinfo  {journal} {Nat. Phys.}\ }\textbf {\bibinfo
  {volume} {15}},\ \bibinfo {pages} {887} (\bibinfo {year} {2019})}\BibitemShut
  {NoStop}%
\bibitem [{\citenamefont {Yupeng~Wang}\ and\ \citenamefont
  {Shi}(2015)}]{offdiagonalBetheAnsatz}%
  \BibitemOpen
  \bibfield  {author} {\bibinfo {author} {\bibfnamefont {J.~C.}\ \bibnamefont
  {Yupeng~Wang}, \bibfnamefont {Wen-Li~Yang}}\ and\ \bibinfo {author}
  {\bibfnamefont {K.}~\bibnamefont {Shi}},\ }\href@noop {} {\emph {\bibinfo
  {title} {Off-Diagonal Bethe Ansatz for Exactly Solvable Models}}}\ (\bibinfo
  {publisher} {Springer, Berlin, Heidelberg},\ \bibinfo {year}
  {2015})\BibitemShut {NoStop}%
\bibitem [{\citenamefont {Lee}\ \emph {et~al.}(2013{\natexlab{b}})\citenamefont
  {Lee}, \citenamefont {Gopalakrishnan},\ and\ \citenamefont
  {Lukin}}]{Lukin2013MeanfieldclosedXYZ}%
  \BibitemOpen
  \bibfield  {author} {\bibinfo {author} {\bibfnamefont {T.~E.}\ \bibnamefont
  {Lee}}, \bibinfo {author} {\bibfnamefont {S.}~\bibnamefont {Gopalakrishnan}},
  \ and\ \bibinfo {author} {\bibfnamefont {M.~D.}\ \bibnamefont {Lukin}},\
  }\bibfield  {title} {\enquote {\bibinfo {title} {Unconventional magnetism via
  optical pumping of interacting spin systems},}\ }\href {\doibase
  10.1103/PhysRevLett.110.257204} {\bibfield  {journal} {\bibinfo  {journal}
  {Phys. Rev. Lett.}\ }\textbf {\bibinfo {volume} {110}},\ \bibinfo {pages}
  {257204} (\bibinfo {year} {2013}{\natexlab{b}})}\BibitemShut {NoStop}%
\bibitem [{\citenamefont {Huybrechts}\ \emph {et~al.}(2020)\citenamefont
  {Huybrechts}, \citenamefont {Minganti}, \citenamefont {Nori}, \citenamefont
  {Wouters},\ and\ \citenamefont {Shammah}}]{Huybrechts2020meanfieldXYZ}%
  \BibitemOpen
  \bibfield  {author} {\bibinfo {author} {\bibfnamefont {D.}~\bibnamefont
  {Huybrechts}}, \bibinfo {author} {\bibfnamefont {F.}~\bibnamefont
  {Minganti}}, \bibinfo {author} {\bibfnamefont {F.}~\bibnamefont {Nori}},
  \bibinfo {author} {\bibfnamefont {M.}~\bibnamefont {Wouters}}, \ and\
  \bibinfo {author} {\bibfnamefont {N.}~\bibnamefont {Shammah}},\ }\bibfield
  {title} {\enquote {\bibinfo {title} {Validity of mean-field theory in a
  dissipative critical system: Liouvillian gap, $\mathbb{PT}$-symmetric
  antigap, and permutational symmetry in the $\mathit{XYZ}$ model},}\ }\href
  {\doibase 10.1103/PhysRevB.101.214302} {\bibfield  {journal} {\bibinfo
  {journal} {Phys. Rev. B}\ }\textbf {\bibinfo {volume} {101}},\ \bibinfo
  {pages} {214302} (\bibinfo {year} {2020})}\BibitemShut {NoStop}%
\end{thebibliography}%

\clearpage
\onecolumngrid
\makeatletter
\setcounter{figure}{0}
\setcounter{equation}{0}
\renewcommand{\thefigure}{S\@arabic\c@figure}
\renewcommand \theequation{S\@arabic\c@equation}
\renewcommand \thetable{S\@arabic\c@table}

\begin{center} 
	{\large \bf Supplementary Material for: Solving the Liouvillian Gap with Artificial Neural Networks}
\end{center}

\section{The stochastic reconfiguration algorithm}
As mentioned in the main text, we adopt the stochastic reconfiguration (SR) method \cite{Sorella2007Weak,Carleo2016Solving,Choo2018Symmetries} to generate the real time evolution of the ansatz density matrix $\rho'$. We note that for closed quantum systems, this kind of variational optimization can be achieved equivalently by the standard stochastic gradient descent (SGD) or imaginary time evolution (via SR). However, due to the non-Hermiticity of Liouvillian superoperator $\mathcal{L}$, the orthogonality of right eigenstates is lost. The extreme value of $\text{Re}(\langle\mathcal{L}\rangle)$, which is usually used as the cost function, generally will not correspond to the Liouvillian gap $\Delta$. The appearance of crossing terms like $\langle\rho_i|\rho_j\rangle\ (\{\ket{\rho_i}\}\text{ are right eigenstates of }\mathcal{L},\ i\neq j)$ will lead to the failure of the former method for open quantum systems. This property can be clearly observed from Fig. 2 of the main text, where during the converging process, $\text{Re}(\avr{\mathcal{L}})$ has 
once exceeded the value of $-\Delta$.

In the framework of SR, given a variational ansatz $\ket{\rho'(\{\alpha_k\})}$, where $\{\alpha_k\}$ stand for the real-number  variational parameters like $\{a_j,b_j,c_k,W_{j,k}^{R(L)}\}$ in the RBM, we need to optimize $\{\alpha_k\}$ such that the trial bi-base wavefunction will finally converge to the first decay modes. First we define the logarithmic derivative operator for each parameter $\alpha_k$ as $O_k$. Note that $O_k$ is diagonal under computational bi-bases $\{\ket{x}=\ket{\sigma_{1,R}^z,\sigma_{2,R}^z,\cdots,\sigma_{N,R}^z,\sigma_{1,L}^z,\sigma_{2,L}^z,\cdots,\sigma_{N,L}^z}\}$:
\begin{equation}
    \bra{x'}O_k\ket{x} = \delta_{x'x}\partial_k \ln \bra{x}\rho'(\{\alpha_k\})\rangle. 
\end{equation}

Consider a small change $\{\delta\alpha_k\}$ on the parameters with respect to the initial values $\{\alpha_k^0\}$
\begin{equation}
    \alpha_k = \alpha_k^0 + \delta \alpha_k.
\end{equation}
The corresponding bi-base wavefunction will deviate from the original term $\ket{\rho'^0}$ by
\begin{equation}
    \ket{\rho'} = \ket{\rho'^0} + \sum_k \delta \alpha_k O_k \ket{\rho'^0}.
\end{equation}

The stochastic reconfiguration scheme proceeds by performing a series of infinitesimal real time evolution governed by the Liouvillian superoperator $\mathcal{L}$, which up to the first order of learning rate $\epsilon$ is given by
\begin{equation}
    \ket{\rho'_{\text{e(xact)}}} = e^{\epsilon\mathcal{L}}\ket{\rho'^0} \approx (1+\epsilon \mathcal{L})\ket{\rho'^0}.
\end{equation}

Now we want to find out the optimal updated parameters $\{\alpha_k\}$ to maximize the overlap between $\ket{\rho'}$ and  $\ket{\rho'_{e}}$. This maximization is equivalent to requiring
\begin{equation}
    \innerproduct{\rho'_e}{\rho'}\innerproduct{\rho'}{\rho'_e} = \innerproduct{\rho'_e}{\rho'_e}\innerproduct{\rho'}{\rho'},
\end{equation}
after some algebraic operations and dropping high-order terms like $\epsilon^2$ or $\epsilon\delta\alpha\delta\alpha$, we obtain the linear equation

\begin{align}
    &\sum_{k'} \left(\avr{O_k^\dagger O_{k'}} + \avr{O_{k'}^\dagger O_{k}} - \avr{O_k^\dagger}\avr{O_{k'}} - \avr{O_{k'}^\dagger}\avr{O_{k}}\right)\delta \alpha_{k'} \nonumber\\
    &= \epsilon\left(\avr{\mathcal{L}^\dagger O_k} + \avr{O_k^\dagger\mathcal{L}} - \avr{\mathcal{L}^\dagger} \avr{O_k} - \avr{O_k^\dagger}\avr{\mathcal{L}}\right)
    \label{linearequation}
\end{align}
where $\langle\cdot\rangle$ denotes the average on the state $\ket{\rho'^0}$.

In short Eq. (\ref{linearequation}) can be written as $\sum_{k'}S_{kk'}\delta\alpha_{k'} = \epsilon F_k$, where $S$ and $F$ are usually called the covariance matrix and force. Finally by solving this linear equation the updated parameter vector can be computed as $\delta\alpha=\epsilon S^{-1}F$. Usually a regularization on $S$ \cite{Carleo2016Solving,Choo2018Symmetries} will be applied to decrease the fluctuation error generated in the Monte Carlo process that will be discussed in the next section:
\begin{equation}
    \widetilde{S} = S + \lambda I\quad \lambda\in[10^{-4},10^{-2}].
     \label{S matrix}
\end{equation}
We continuously repeat the iteration above to update the parameters until convergence.

Besides, for the case (iii) of first decay modes mentioned in the main text, in order to make the RBM ansatz further converge to the first decay mode with the minimal imaginary part, we should add another imaginary time evolution for $\rho'$ under $i\mathcal{L}$ with smaller learning rate $\beta\epsilon$. Hence, another force $F'$ will be added
\begin{equation}
    F_k' = -i\avr{\mathcal{L}^\dagger O_k} + i\avr{O_k^\dagger\mathcal{L}} + i\avr{\mathcal{L}^\dagger} \avr{O_k} - i\avr{O_k^\dagger}\avr{\mathcal{L}},
\end{equation}
and the linear equation becomes
\begin{equation}
    \sum_{k'}S_{kk'}\delta\alpha_{k'} = \epsilon (F_k + \beta F_k')\quad \beta\in[10^{-3},10^{-2}].
\end{equation}

Under the joint evolution of $\mathcal{L}+i\beta\mathcal{L}$,
\begin{align}
    \rho' &= e^{(\mathcal{L}+i\beta\mathcal{L})t}\rho'^0\nonumber\\
    &= c_0\rho_0 + \sum_{i\neq 0} c_i e^{(\lambda_i+i\beta\lambda_i)t}\rho_i \quad(c_0=0)\nonumber\\
    &= \sum_{i\neq 0}  c_i e^{(\text{Re}(\lambda_i)-\beta \text{Im}(\lambda_i))t}e^{i(\text{Im}(\lambda_i)+\beta \text{Re}(\lambda_i))t}\rho_i,
\end{align}
where $\rho_0$ denotes the steady state and $\{\rho_i\}_{i\neq 0}$ are other decay eigenmodes. After long enough time, clearly $\rho'$ will converge to the first decay mode with the minimal imaginary part. A numerical example belonging to this case will be shown in the last section.

\section{The Markov chain Monte Carlo}
In order to efficiently compute the average of observables mentioned in the previous section, we introduce the standard Markov chain Monte Carlo (MCMC) method as follows. Given an observable $\hat{B}$ (like $O_k$ and $\mathcal{L}$), we convert its average into the stochastic form:
\begin{align}
    \avr{\hat{B}} &= \frac{\bra{\rho'} \hat{B} \ket{\rho'}}{\langle\rho' |\rho'\rangle} \nonumber\\
    &= \sum_{x,x'} \frac{\langle\rho'\ket{x}\bra{x} \hat{B}\ket{x'}\bra{x'}\rho'\rangle}{\langle\rho'|\rho'\rangle} \nonumber\\
    &= \sum_{x,x'} \frac{\langle\rho'\ket{x}\bra{x}\rho'\rangle}{\langle\rho' |\rho'\rangle}
    \frac{\bra{x}\hat{B}\ket{x'}\bra{x'}\rho'\rangle}{\bra{x}\rho'\rangle} \nonumber\\
    &=\mathbb{E}\left(\sum_{x'}\frac{\bra{x}\hat{B}\ket{x'}\bra{x'}\rho'\rangle}{\bra{x}\rho'\rangle}\right)\quad\text{by sampling a distribution } |\bra{x}\rho'\rangle|^2.
\end{align}

Since $O_k$ is diagonal under computational bi-bases, the explicit expressions of $O_k$ for the RBM ansatz  can be directly read out as follows:
\begin{equation}
    O_{\text{Re}(a_j)} = \sigma_{j,R}^z \quad O_{\text{Im}(a_j)} = i\sigma_{j,R}^z \quad O_{\text{Re}(b_j)} = \sigma_{j,L}^z \quad O_{\text{Im}(b_j)} = i\sigma_{j,L}^z,
\end{equation}
\begin{equation}
    O_{\text{Re}(c_k)} = \tanh{X_k} \quad O_{\text{Im}(c_k)} = i\tanh{X_k},
\end{equation}
\begin{equation}
    O_{\text{Re}(W_{k,j}^{R(L)})} = \sigma_{j,R(L)}^z\tanh{X_k} \quad O_{\text{Im}(W_{k,j}^{R(L)})} = i\sigma_{j,R(L)}^z\tanh{X_k}.
\end{equation}

For achieving the stochastic average, we will generate a Markov chain of computational bi-bases $x_1\rightarrow x_2\rightarrow x_3\rightarrow \cdots\rightarrow x_{N_s}$ with total length $N_s$ by sampling $|\rho'|^2$. This process can be realized by the well-known Metropolis-Hasting algorithm \cite{metropolis1953equation}:
At step $i$, we randomly flip 1 to 4 spins in $x_i$ to obtain a new sample and calculate the following acceptance probability to determine whether to accept it:
\begin{equation}
    A(x_i\rightarrow x_{i+1}) = \min\left(1, \left|\frac{\rho_{x_{i+1}}}{\rho_{x_{i}}} \right|^2 \right).
\end{equation}
Then we use the ensemble observable average of the Markov chain to estimate the stochastic average. Besides, considering the thermalization process, we need to drop the first $5\%$ of the Markov chain.

After each SR iteration, we should recompute the trace of updated RBM ansatz $\rho_\text{RBM}$ in order to make sure $\tr{\rho'}=0$, as mentioned in the main text. This step can be achieved directly in small system size (typically $N \leq 10$) and also can be implemented by a sampling paradigm for larger system size as follows.
\begin{align}
    \frac{\tr{\rho_\text{RBM}}}{\tr{\rho_0'}} &= \frac{\bra{I}\rho_\text{RBM}\rangle}{\bra{I}\rho_0'\rangle}\qquad(\ket{I} = \sum_{l=0}^{2^N-1} \ket{l,l} \text{ which is the bi-base state corresponding to the identity matrix})\nonumber\\
    &=\sum_{l=0}^{2^N-1}\frac{\bra{I}l,l\rangle\bra{l,l}\rho_\text{RBM}\rangle}{\bra{I}\rho_0'\rangle}\nonumber\\
    &=\sum_{l=0}^{2^N-1}\frac{\bra{I}l,l\rangle\bra{l,l}\rho_0'\rangle}{\bra{I}\rho_0'\rangle}\frac{\bra{l,l}\rho_\text{RBM}\rangle}{\bra{l,l}\rho_0'\rangle}\nonumber\\
    &=\mathbb{E}\left(\frac{\bra{l,l}\rho_\text{RBM}\rangle}{\bra{l,l}\rho_0'\rangle}\right)\quad\text{by sampling a distribution } p(l)\propto \bra{l,l}\rho_0'\rangle \text{ on }l \ (l = 0,1\cdots,2^N-1)
\end{align}
The diagonal real ancillary $\rho_0'$ should be chosen appropriately according to the specific Liouvillian $\mathcal{L}$, mainly resembling its steady state property. For example, $\rho_0'$ for the 1D/2D XXZ model corresponds to the bi-base state with all spins pointing down, so that we can impose $\rho_\text{RBM}(0,0)=\tr{\rho_\text{RBM}}$, whereas $\rho_0'$ used for the XYZ model is the identity matrix. Besides, in order to obtain better convergence performance, the derivative of $\tr{\rho_\text{RBM}}$ can also be taken into consideration similarly.

\begin{align}
    & \partial_k \frac{\tr{\rho_\text{RBM}}}{\tr{\rho_0'}} = \partial_k\frac{\bra{I}\rho_\text{RBM}\rangle}{\bra{I}\rho_0'\rangle}=\mathbb{E}\left(O_k(l,l)\frac{\bra{l,l}\rho_\text{RBM}\rangle }{\bra{l,l}\rho_0'\rangle}\right)\nonumber\\
    & \text{by sampling a distribution } p(l)\propto \bra{l,l}\rho_0'\rangle \text{ on }l \ (l = 0,1\cdots,2^N-1)
\end{align}

In our numerical calculations, the typical sampling size $N_s$ is 2 to 5 times as many as the variational parameters. The ratio between hidden neurons and visible neurons $M/2N$ is around 3 to 6. The learning rate is given by $\epsilon=\max(0.01, 0.1\times 0.96^p),\beta = 0.005\sim 0.05$ ($p$ is the iteration step). The regularization parameter takes $\lambda=\max(10^{-4},0.9^p)$. 
 
\section{Representation power of the RBM and computational cost}
 In this section, in order to make the article more self-contained and accessible to the general audience, we will provide a brief introduction for the representation power of the RBM ansatz and discuss about the computational cost. For an existing review of the RBM method, readers can refer to \cite{melko2019restrictedreview}.

As mentioned in the main text, owing to the structure flexibility and long-range connections of neural networks,the restricted Boltzmann machine (RBM) ansatz is able to represent quantum states with even volume-law entanglement in high dimensional systems \cite{Deng2017Quantum}. The cases with volume-law entanglement, as expected, will appear more frequently in the open quantum systems. In contrast to the DMRG (density-matrix-renormalization-group) or other tensor-network approaches, entanglement is not a limiting factor for our RBM approach. 
Yet, it is also important to clarify that the RBM ansatz cannot represent all the volume-law entangled states efficiently. Indeed, it has been rigorously proven in Ref. \cite{Gao2017Efficient} that there exist certain quantum states (which are generated either by polynomial-size quantum circuits or as ground states of certain gapped Hamiltonians) that cannot be represented by RBMs with a polynomial number of parameters,  unless the polynomial
hierarchy (a generalization of the famous P versus NP problem in computer science) collapses (which is widely believed to be unlikely).

Finding out the key properties of the Liouvillian superoperators that warrant the effectiveness of the RBM approach is of both fundamental and practical importance. However, this may require new physical concepts and a deeper understanding of artificial neural networks, similar to the case of how we understand the effectiveness
of the DMRG algorithm from the entanglement perspective \cite{Schollwock2005TheDMRG}. There are still many open questions about the representation power and effectiveness of the RBM method, which need further explorations in the future.

We also remark that our RBM approach may carry over straightforwardly to systems with higher spins, interacting bosons and fermions, or models with long-range interactions. In fact, for closed quantum systems the RBM method has been explored for  bosons \cite{Saito2017Solving, Choo2018Symmetries}, fermions \cite{Nomura2017Restricted,Choo2020Fermionic}, and models with long-range interactions  \cite{Deng2017Quantum}.  
 The general ideas are as follows. For higher spins or bosonic systems, we need to increase the degrees of freedom for the neurons in the RBM representation, whereas for fermionic systems, a Slater determinant or a pair-product wave function can be added to satisfy the Fermi-Dirac statistics. 
To demonstrate the effectiveness of our RBM approach in dealing with long-range interactions, we have also carried out numerical calculations for the Liouvillian gap of the dissipative long-range XXZ model and our results are plotted in Fig. \ref{figure: SMfig3}.
 

As for the computational cost of the RBM method, the total complexity for each SR iteration is bounded by the MCMC sampling and the matrix inversion operation. For the first part, in the numerical calculation we usually take the Monte Carlo sampling size proportional to the number of variational parameters, $O(N_\text{par})$. According to Eqn. \ref{linearequation}, the operation times to generate one covariance matrix $S$ are $O(N_\text{par}^3)$. However, the sampling step can be significantly sped up by parallel processing (on CPU or GPU) and batch processing. Concretely, we do not update all the parameters in one SR iteration, but divide them up and update separately in parallel. For the matrix inversion part, $\delta\alpha=\epsilon S^{-1}F$, direct Gauss elimination will be bounded by the complexity of $O(N_\text{par}^3)$. Yet, following the method mentioned in the Supplementary Material of \cite{Carleo2016Solving}, this cost can be reduced to $O(N_\text{par}^2)$ by means of iterative solvers which never form the covariance matrix $S$ explicitly. Finally, the number of parameters $N_\text{par}$ depends on the specific structure of RBM ansatz. In most  cases, each hidden neuron is coupled to all the visible neurons and the number of hidden neurons is usually on the same order of that for visible neurons, $M\sim N$, hence $N_\text{par} = O(M\times N)=O(N^2)$. This number of variational parameters may be further reduced by exploiting the translational symmetry (see the last section for details) or the locality property (one hidden neuron only couples to nearest $O(1)$ visible neurons), which will lead to significant speed up in practice. 


In summary, the total computational complexity for each SR iteration is upper bounded by a polynomial cost $O(N_\text{par}^3)$, and many numerical techniques mentioned above can significantly reduce this complexity. In the generalization of RBM method to bosonic systems, fermionic systems or lattice models with long-range interactions, the total computational cost may increase a little bit due to the increase of the degrees of freedom of the neurons or the extra Slater determinant.  However, in all the cases the complexity still scales polynomially with the system size $N$ and many numerical tricks have been developed and adopted to speed up the computation process.

\section{Eigenstates of the dissipative XXZ model}
As mentioned in the main text, the Liouvillian spectrum of dissipative XXZ model coincides with that of $\tilde{\mathcal{L}}'=H_R+H_L$, but not for the eigenstates. In this section we will give the general conditions for this kind of coincidence, and demonstrate how to construct the eigenstates of $\mathcal{L}$ from that of $\tilde{\mathcal{L}}'$ by following calculations in \cite{torre2014closedform}.

A Liouvillian superoperator $\mathcal{L}$, in the representation of Choi-Jamio\l kowski isomorphism \cite{Zwolak2004Mixed,kshetrimayum2017simpleTNsteady} can always be decomposed into three parts:
\begin{equation}
\tilde{L}=(-iH-\sum_\mu L_\mu^\dagger L_\mu)\otimes I+I\otimes(iH^T-\sum_\mu L_\mu^T L_\mu^*)+\sum_\mu 2L_\mu\otimes L_\mu^*=H_\text{NH}\otimes I+I\otimes H_\text{NH}^*+\sum_{\mu}D_\mu,
\end{equation}
where $H_\text{NH}=-iH-\sum_\mu L_\mu^\dagger L_\mu$ is the effective non-Hermitian Hamiltonian which governs the short-time coherent dynamics, and $D_\mu=2L_\mu\otimes L_\mu^*$ is the decoherence term for the quantum jump channel $\mu$. Consider the case when spectrum of $H_\text{NH}$ can be solved exactly, which is easier than the whole Liouvillian spectrum in general since the dimension is reduced from $2^{2N}$ to $2^N$, and there exists a conserved physical quantity $M$ which commutes with the effective Hamiltonian:
\begin{equation}
[M,H_\text{NH}]=0.
\end{equation}
Meanwhile $L_\mu$ is the lowering operator of $M$:
\begin{equation}
[M,L_\mu]=-m_\mu L_\mu
\end{equation}
with positive real number $m_\mu>0$. 

According to the assumptions, the simultaneous eigenstates of $H_\text{NH}$ and $M$ have been solved, and denoted by two indices $|m,j\rangle$:
\begin{equation}
H_\text{NH}|m,j\rangle=E_{m,j}|m,j\rangle, M|m,j\rangle=m|m,j\rangle.
\end{equation}
The corresponding left eigenstates are $|\overline{m,j}\rangle$ with $H_\text{NH}^\dagger|\overline{m,j}\rangle=E_{m,j}^*|\overline{m,j}\rangle$. For simplicity we assume $m$ to be integers, the lowest value of $m$ is zero, and $m_\mu=1$ for all channels $\mu$. Index $j$ labels different eigenstates with the same $m$, so the range of $j$ depends on $m$, denoted by $d_m$ in the following text. The effect of $L_\mu$ is to lower $m$ by $1$:
\begin{equation}
L_\mu|m,j\rangle=\sum_k^{d_{m-1}}l_{\mu,jk}^{(m)}|m-1,k\rangle,
\end{equation}
here $l_{\mu,jk}^{(m)}=\langle\overline{m-1,k}|L_\mu|m,j\rangle$.

For the coherent part of Liouvillian superoperator $\tilde{\mathcal{L}}'=H_\text{NH}\otimes I+I\otimes H_\text{NH}^*$, we can construct its eigenstates by the direct product:
\begin{equation}
\tilde{\mathcal{L}}'|m,j\rangle\otimes|n,k^*\rangle=(E_{m,j}+E_{n,k}^*)|m,j\rangle\otimes|n,k^*\rangle,
\end{equation}
the superscript $*$ on the state $|n,k^*\rangle$ means the conjugation of wavefunctions. Sort the eigenstates by $m+n$ in increasing order, then due to the pure loss property of $D_\mu$,
\begin{equation}
D_\mu |m,j\rangle\otimes|n,k^*\rangle=\sum_{j'}^{d_{m-1}}\sum_{k'}^{d_{n-1}}2l_{\mu,jj'}^{(m)}l_{\mu,kk'}^{(n)*}|m-1,j'\rangle\otimes|n-1,k'^*\rangle.
\end{equation}
The decoherent term takes an upper triangular form in this set of bases, so $\tilde{\mathcal{L}}'$ and $\mathcal{L}$ share the same spectrum.

As for the eigenstates of $\mathcal{L}$, the one with eigenvalue $\lambda_{m,j;n,k}=E_{m,j}+E_{n,k}^*$ can be constructed as the linear superposition of $|m,j\rangle\otimes|n,k^*\rangle$ together with all the other states having smaller $m$ and $n$. For concreteness, given $m\le n$ the eigenstate can be expanded as
\begin{equation}
|m,j;n,k\rangle=|m,j\rangle\otimes|n,k^*\rangle+\sum_{r=1}^m\sum_{j'}^{d_{m-r}}\sum_{k'}^{d_{n-r}}C_{jk;j'k'}^{(mn),r} |m-r,j'\rangle\otimes|n-r,k'^*\rangle.
\end{equation}
Then we solve the set of coefficients $C_{jk;j'k'}^{(mn),r}$ by the eigenvalue equation
\begin{equation}
\tilde{\mathcal{L}}|m,j;n,k\rangle=\lambda_{m,j;n,k}|m,j;n,k\rangle.
\end{equation}
It turns out that the equations are iterative, so that the algebraic relation from $r-1$ to $r$ is given by
\begin{equation}
C_{jk;j'k'}^{(mn),r}(\lambda_{m,j;n,k}-\lambda_{m-r,j';n-r,k'})=\sum_{j''}^{d_{m-r+1}}\sum_{k''}^{d_{n-r+1}} C_{jk;j''k''}^{(mn),r-1}2\sum_\mu l_{\mu,j''j'}^{(m-r+1)} l_{\mu,k''k'}^{(n-r+1)*}.
\end{equation}
For example, given $C_{jk;j'k'}^{(mn),0}=\delta_{jj'}\delta_{kk'}$, from $r=0$ to $r=1$
\begin{equation}
C_{jk;j'k'}^{(mn),1}=\frac{2}{\lambda_{m,j;n,k}-\lambda_{m-1,j';n-1,k'}}\sum_\mu l_{\mu,jj'}^{(m)}l_{\mu,kk'}^{(n)*}.
\end{equation}
When $\lambda_{m,j'n,k}=\lambda_{m-1,j';n-1,k'}$, it implies that the Liouvillian superoperator is tuned to an exceptional point where both the eigenvalues and the corresponding eigenstates coincide, characterizing a unique feature of non-Hermitian matrices.

In the dissipative XXZ model, the right (left) single-magnon excitation $|1,j\rangle\otimes|0\rangle (|0\rangle\otimes|1,j^*\rangle)$, which decides the Liouvillian gap, has no matrix elements for $D_\mu$ since $l^{(0)}_{\mu,jk}=0$, so the first decay modes also coincide with single-magnon eigenstates of $\tilde{\mathcal{L}}'$.

\section{The Bethe ansatz solution for the effective Hamiltonian}
In the main text we claim that the eigenspectrum of the Liouvillian superoperator $\mathcal{L}$ for the dissipative XXZ model can be exactly obtained in 1D, by solving the effective non-Hermitian Hamiltonian. In this section we will demonstrate how to apply the Bethe ansatz on the effective Hamiltonian, for which eigenstates are magnon excitations over the reference state. Next we derive the Bethe equations to determine the quasi-momentum for two-magnon excitations and generalize the results to multi-magnon cases.

Take the right Hamiltonian in the main text
\begin{equation}
    H_R=\sum_{i=1}^{N}[-i J(S_{i,R}^+S_{i+1,R}^-+S_{i,R}^-S_{i+1,R}^+)-i J_z S_{i,R}^z S_{i+1,R}^z-\frac{\gamma}{2}(S_{i,R}^z+\frac{1}{2})].
\end{equation}
The periodic boundary condition is imposed by assuming $S_{N+1}=S_1$. The state with all $S_{i,R}^z=-\frac{1}{2}$ is the eigenstate of $H_R$ with eigenvalue $E_g=-iJ_z\frac{N}{4}$, and will be chosen as the reference state.

Following the well-known Bethe ansatz, the eigenstates are magnon excitations which mean spin flips from down to up, creating spin-one quasi-particles. Consider single-magnon excitations firstly:
\begin{equation}
    |\Psi_1\rangle=\sum_{j=1}^{N}e^{ikj}|j\rangle,
\end{equation}
where $|j\rangle$ denotes the state with only the $j^{\text{th}}$ spin being flipped up and others remaining down. To fulfill the boundary condition we need $e^{ikN}=1$ so that the quasi-momentum $k$ can only take quantized values $k=\frac{2\pi n}{N}, n=0,\cdots,N-1$. The eigenvalue of the state is $E_{k}=-\frac{\gamma}{2}-i(2J \text{cos}k-J_z)+E_g$, whose real part gives the Liouvillian gap.

As for the two-magnon cases, considering the exchange of two magnons, the state is given by
\begin{equation}
     |\Psi_2\rangle=\sum_{j_1<j_2}(c_1 e^{i(k_1j_1+k_2j_2)}+c_2 e^{i(k_1j_2+k_2j_1)})|j_1,j_2\rangle.
\end{equation}
This is the eigenstate with eigenenergy $E_{k_1,k_2}=-\gamma-i(2J \text{cos}k_1+2J \text{cos}k_2-2J_z)+E_g$ if and only if the two coefficients fulfill 
\begin{equation}
    \frac{c_1}{c_2}=-\frac{J(e^{i(k_{1}+k_{2})}+1)-J_{z}e^{ik_{1}}}{J(e^{i(k_{1}+k_{2})}+1)-J_{z}e^{ik_{2}}}.
\end{equation}
Moreover, by imposing the boundary condition $c_1/c_2=e^{ik_1N}=e^{-ik_2N}$, we obtain Bethe equations to determine the possible discrete values of quasi-momentum:
\begin{equation}
    e^{ik_1N}=e^{-ik_2N}=-\frac{J(e^{i(k_{1}+k_{2})}+1)-J_{z}e^{ik_{1}}}{J(e^{i(k_{1}+k_{2})}+1)-J_{z}e^{ik_{2}}}.
\end{equation}
The framework above can be generalized to the $m$-magnon wavefunction:
\begin{equation}
|\Psi_m\rangle=\sum_{j_1<j_2<\cdots<j_m}[(\sum_{\mathcal{P}}c_{\mathcal{P}} e^{i\sum_n^m k_{\mathcal{P}n} j_n})|j_1,j_2,\cdots,j_m\rangle],
\end{equation}
where all possible permutations $\mathcal{P}$ of integers $1,\cdots,m$ are taken into the summation. Now the eigenvalues are $E({\{k_j\}_m})=-\frac{\gamma}{2}m-i\sum_{j=1}^m (2J\text{cos}k_j-J_z)+E_g$, while Bethe equations become
\begin{equation}
    e^{ik_{j}N}=\prod_{l\ne j}-\frac{J(e^{i(k_{j}+k_{l})}+1)-J_{z}e^{ik_{j}}}{J(e^{i(k_{j}+k_{l})}+1)-J_{z}e^{ik_{l}}}.
\end{equation}
For more details about Bethe ansatz, readers can refer to Ref. \cite{offdiagonalBetheAnsatz}.

\section{The mean-field theory of dissipative XYZ model}
When the spin-spin interaction is anisotropic, the U(1) symmetry is broken, so that the Liouvillian spectrum is no longer exactly solvable. In this section we will apply a mean-field approximation to obtain the expectation value of physical quantities of the steady state, and identify the second-order phase transition, accompanied by the closing of Liouvillian gap. In one-dimensional system the mean-field theory fails due to strong quantum fluctuations, which are manifested by numerical results that the gap of one phase is much smaller than the other but never approaches zero.

To analyze the evolution of expectation value of operators, the adjoint Lindblad equation is needed:
\begin{equation}
\mathcal{L}^a(O)\equiv\frac{d O}{dt}=i[H,O]+\sum_\mu 2L_\mu^\dagger OL_\mu-\{L_\mu^\dagger L_\mu,O\}.
\end{equation}
Analogous to its counterpart ``Heisenberg picture'' in closed quantum systems, the time-dependent operator is defined to preserve the expectation value in the original picture:
\begin{equation}
\text{Tr}(O(t)\rho(0))=\text{Tr}(O\rho(t)).
\end{equation}

With the adjoint Lindblad master equation, the evolution for local spin operators $S_i^\alpha,\alpha=x,y,z$ in the dissipative XYZ model are
\begin{eqnarray}\label{adjoint}
\frac{dS^x_i}{dt}&=&J_y(S_{i-1}^y+S_{i+1}^y)S_i^z-J_z(S_{i-1}^z+S_{i+1}^z)S_i^y-\frac{\gamma}{2}S_i^x,\nonumber\\
\frac{dS^y_i}{dt}&=&J_z(S_{i-1}^z+S_{i+1}^z)S_i^x-J_x(S_{i-1}^x+S_{i+1}^x)S_i^z-\frac{\gamma}{2}S_i^y,\nonumber\\
\frac{dS^z_i}{dt}&=&J_x(S_{i-1}^x+S_{i+1}^x)S_i^y-J_y(S_{i-1}^y+S_{i+1}^y)S_i^x-\gamma (S_i^z+\frac{1}{2}).
\end{eqnarray}
Assuming the mean-field approximation which states that the many-body density matrix is the tensor product of identical density matrices for each site $\rho=\otimes_{i=1}^N \rho_i$, for every site $i$, $\text{Tr}(S_i^\alpha \rho)=\text{Tr}(S_i^\alpha \rho_i)$ is the same, defined as $\langle S^\alpha\rangle$. While for the operator product,
\begin{equation}
    \text{Tr}(S_i^\alpha S_j^\beta\rho)=\text{Tr}(S_i^\alpha\rho_i)\text{Tr}(S_j^\alpha\rho_j)=\langle S^\alpha\rangle\langle S^\beta\rangle, i\ne j.
\end{equation}

Take the density matrix average of Eq. (\ref{adjoint}). With the above approximation they are reduced to 
\begin{eqnarray}
\frac{d\langle S^x\rangle}{dt}&=&2(J_y-J_z)\langle S^y\rangle \langle S^z\rangle-\frac{\gamma}{2}\langle S^x\rangle,\nonumber\\
\frac{d\langle S^y\rangle}{dt}&=&2(J_z-J_x)\langle S^z\rangle \langle S^x\rangle-\frac{\gamma}{2}\langle S^y\rangle,\nonumber\\
\frac{d\langle S^z\rangle}{dt}&=&2(J_x-J_y)\langle S^x\rangle \langle S^y\rangle-\gamma(\langle S^z\rangle+\frac{1}{2}).
\end{eqnarray}.

For the steady state, we have the expectation value of $S^z$:
\begin{equation}
\langle S^z\rangle=-\frac{\gamma}{4\sqrt{(J_y-J_z)(J_z-J_x)}}.
\end{equation}
The inequality relation $|\langle S^z\rangle|\le 1/2$ must be fulfilled on the steady state, which gives
\begin{equation}
 \gamma^2 < 4(J_y-J_z)(J_z-J_x).
\end{equation}
If so, the polarization on the steady state will deviate from $z$-direction so that $\langle S^x\rangle, \langle S^y\rangle\ne 0$. Moreover, since the parity operator $P=\prod_{j=0}^{N-1}e^{i\pi (S_j^z+\frac{1}{2})}$ commutes with the Liouvillian superoperator, which reverses $S^x$ and $S^y$, steady states must be at least two-fold degenerate and span a steady subspace as the eigenspace of $P$, implying the closing of Liouvillian gap. It is the counterpart of spontaneous symmetry breaking in quantum mechanics, though here the relations between symmetry and conservation laws are more sophisticated than unitary cases. On the other hand, when the parameters violate the inequality, the tensor product ansatz $\rho=\otimes_{
i=1}^N\rho_i$ will fail. The steady state polarization approaches the maximal value $1/2$, and the Liouvillian gap is opened between the unique steady state and the first decay modes. In Fig. 3 of the main text, the system with chosen parameters lies in the degenerate phase, though quantum fluctuations open the gap, while for higher dimensions \cite{Lukin2013MeanfieldclosedXYZ,Jin2016Cluster} or all-to-all connected lattices \cite{Huybrechts2020meanfieldXYZ} the critical dynamics will occur.

\section{More numerical results and discussions}
\begin{figure}
\hspace*{-1\textwidth}
\centering 
\includegraphics[width=1\textwidth]{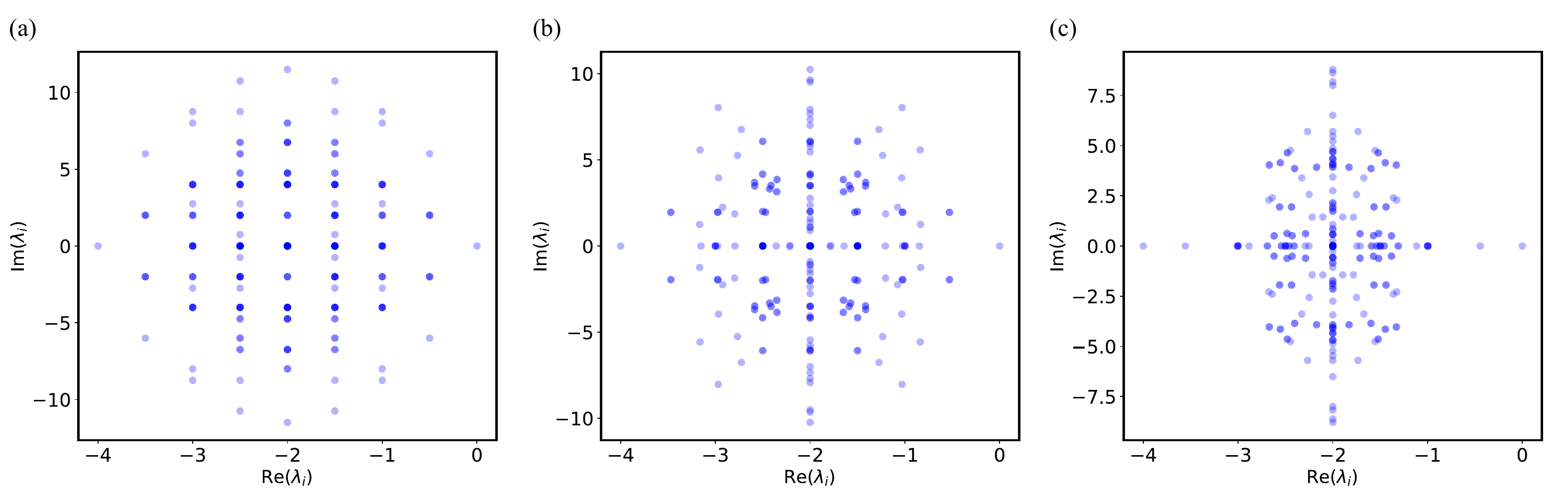}
\caption{The Liouvillian spectrum for the dissipative XYZ model in 1D, obtained by ED. The model parameters are chosen as $N=4$, $\gamma=1$, $J_x=4$, and $J_z=2$. (a) $J_y=4$. The XXZ case with multiple orthogonal first decay modes. (b)  $J_y=3$. The gapped XYZ case with multiple non-orthogonal first decay modes. (c) $J_y=0.5$. The ``gapless" XYZ case with only one first decay mode. The color shade stands for the degeneracy of each eigenvalue $\lambda_i$.
} \label{figure: SMfig1}
\end{figure}

In this section we will provide more numerical results and relevant discussions. In Fig. \ref{figure: SMfig1}, we display the Liouvillian spectrum of 1D dissipative XYZ model with different parameters obtained by exact diagonaliztion (ED). The panel (a), (b), (c) respectively correspond to the XXZ case, gapped XYZ phase, and ``gapless" XYZ phase (actually gapped due to strong quantum fluctuations in 1D). The latter two have been discussed in the previous section. As mentioned in the main text, in comparison with the XXZ case, the reason for the slower convergence of XYZ model is that: The XXZ model has multiple orthogonal first decay modes (both deduced from the third section and tested numerically), like the case (a), so that $\rho'$ only needs to converge to a subspace spanned by these modes. Whereas for the XYZ model, there exists either only one or multiple but non-orthogonal first decay modes (tested numerically), like the case (b) and (c), such that in order to obtain the Liouvillian gap $\Delta$ accurately, $\rho'$ needs to converge to a single first decay mode, which demands extra iteration steps, longer Markov chains and more hidden neurons of the RBM.

\begin{figure}
\hspace*{-0.8\textwidth}
\centering 
\includegraphics[width=0.8\textwidth]{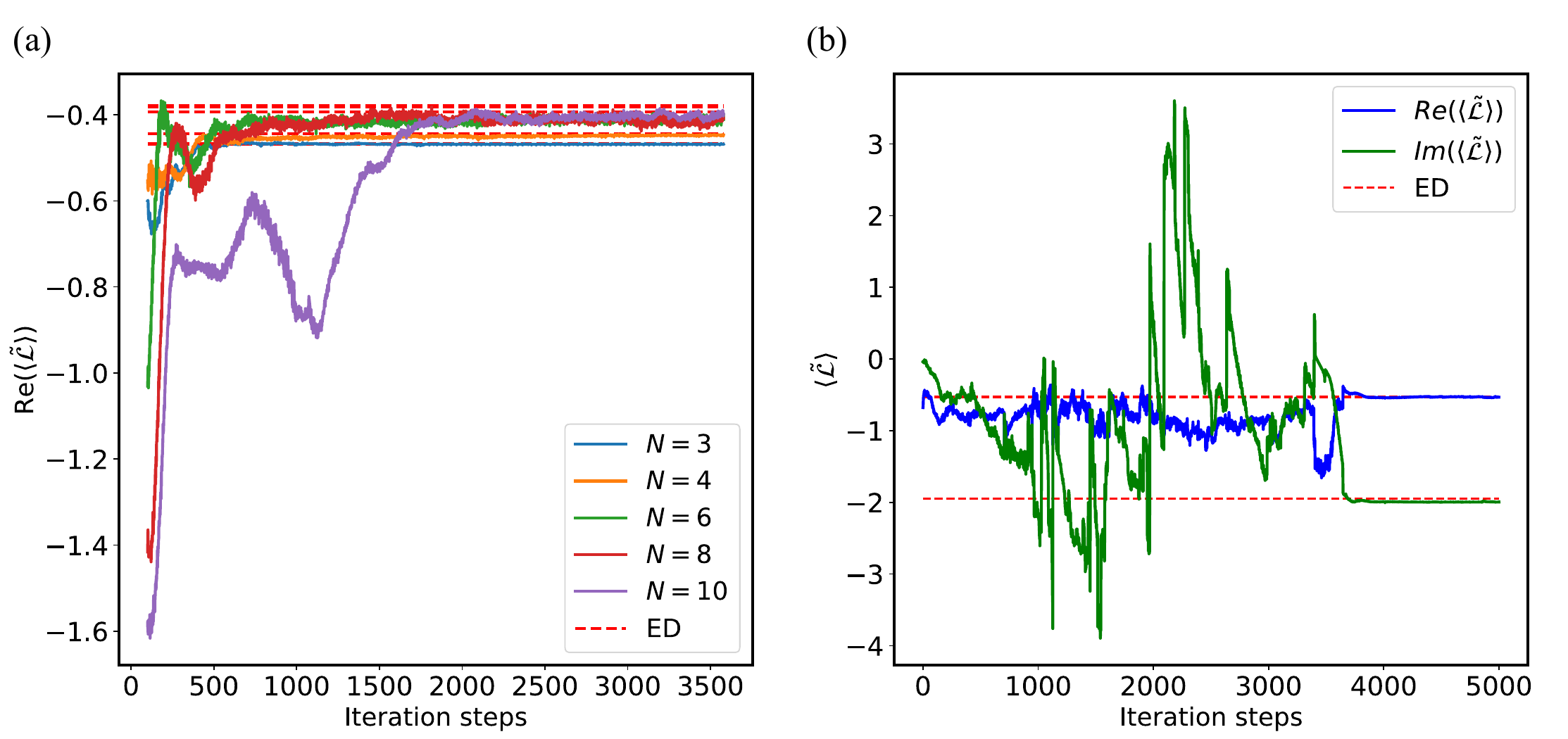}
\caption{The convergence for the Liouvillian gap  of the 1D dissipative XYZ model, obtained by the RBM approach. (a) $\text{Re}\avr{\tilde{\mathcal{L}}}$ as a function of the iteration steps for different lattice size $N$. $J_x=4$, $J_y= 0.5$, $J_z=2$, and $\gamma=1$. The systems lie in the ``gapless" XYZ phase with only one first decay mode. (b) $\text{Re}\avr{\tilde{\mathcal{L}}}$ and $\text{Im}\avr{\tilde{\mathcal{L}}}$ as a function of the iteration steps for the gapped XYZ phase. $N=4$, $J_x=4$, $J_y= 3$, $J_z=2$ and $\gamma=1$. There exist multiple non-orthogonal first decay modes so that we need the joint evolution $\mathcal{L}+i\beta\mathcal{L}$ mentioned in the first section. The ancillary $\rho'_0$ used here corresponds to the bi-base state with all spins pointing down.
} \label{figure: SMfig2}
\end{figure}

For the panel (a) of Fig. \ref{figure: SMfig2}, we show the precise convergence behaviour for Liouvillian gap computation of 1D dissipative XYZ model (``gapless" phase) obtained by the RBM, corresponding to Fig. 3 of the main text. It can be observed that the differences between Liouvillian gap of $N=6,8,10$ are relatively small. The largest relative error is of order $10^{-2}$. The panel (b) shows an example of the gapped XYZ phase with multiple non-orthogonal first decay modes, where we need the joint evolution $\mathcal{L}+i\beta\mathcal{L}$ mentioned in the first section. Due to the smaller learning rate $\beta\epsilon$ on the imaginary part, we need to devote more numerical efforts. The violent oscillation of $\text{Im}\avr{\mathcal{L}}$ before convergence is reasonable since the Liouvillian spectrum is symmetric with respect to the real axis.

For Fig. 2 of the main text, the results of $N = 5 \times 5$ 2D XXZ case are obtained by imposing translational symmetry to the RBM ansatz (actually this case can also be computed directly with longer running time) and its early convergence steps are cut off to fit the plotting range. We choose the ancillary $\rho'_0$ as the bi-base state with all spins pointing down and the translational symmetry for first decay modes can be deduced from the analytical derivations above. Hence, the variational parameters are restricted to be:
\begin{equation}
    a_j = a, \quad 
    b_j = b, 
\end{equation}
\begin{equation}
    c_{1+2qN} = c_{2+2qN} = \cdots = c_{N+2qN},
\end{equation}
\begin{equation}
    c_{1+N+2qN} = c_{2+N+2qN} = \cdots = c_{2N+2qN}, 
\end{equation}
\begin{equation}
    W_{1+2qN,j}^{R(L)} = W_{\sigma^{j-1}(1)+2qN,1}^{R(L)} \quad
    W_{2+2qN,j}^{R(L)} = W_{\sigma^{j-1}(2)+2qN,1}^{R(L)}
    \ 
    \cdots 
    \  W_{N+2qN,j}^{R(L)} = W_{\sigma^{j-1}(N)+2qN,1}^{R(L)}, 
\end{equation}
\begin{equation}
    W_{1+N+2qN,j}^{R(L)} = W_{\sigma^{j-1}(1)+N+2qN,1}^{R(L)} \quad
    W_{2+N+2qN,j}^{R(L)} = W_{\sigma^{j-1}(2)+N+2qN,1}^{R(L)} \ 
    \cdots\  W_{2N+2qN,j}^{R(L)} = W_{\sigma^{j-1}(N)+N+2qN,1}^{R(L)},
\end{equation}
where $j\in\{1,2,\cdots,N\},q\in\{0,1,2,\cdots,M/2N-1\}$, $\sigma$ is the following permutation
\begin{equation}
\sigma=
\left(\begin{array}{ccccccc}
1 & 2 & 3 & 4 & \cdots & N-1 & N\\
N & 1 & 2 & 3 & \cdots & N-2 & N-1
\end{array} 
\right).
\end{equation}

\begin{figure}
\hspace*{-0.4\textwidth}
\centering 
\includegraphics[width=0.4\textwidth]{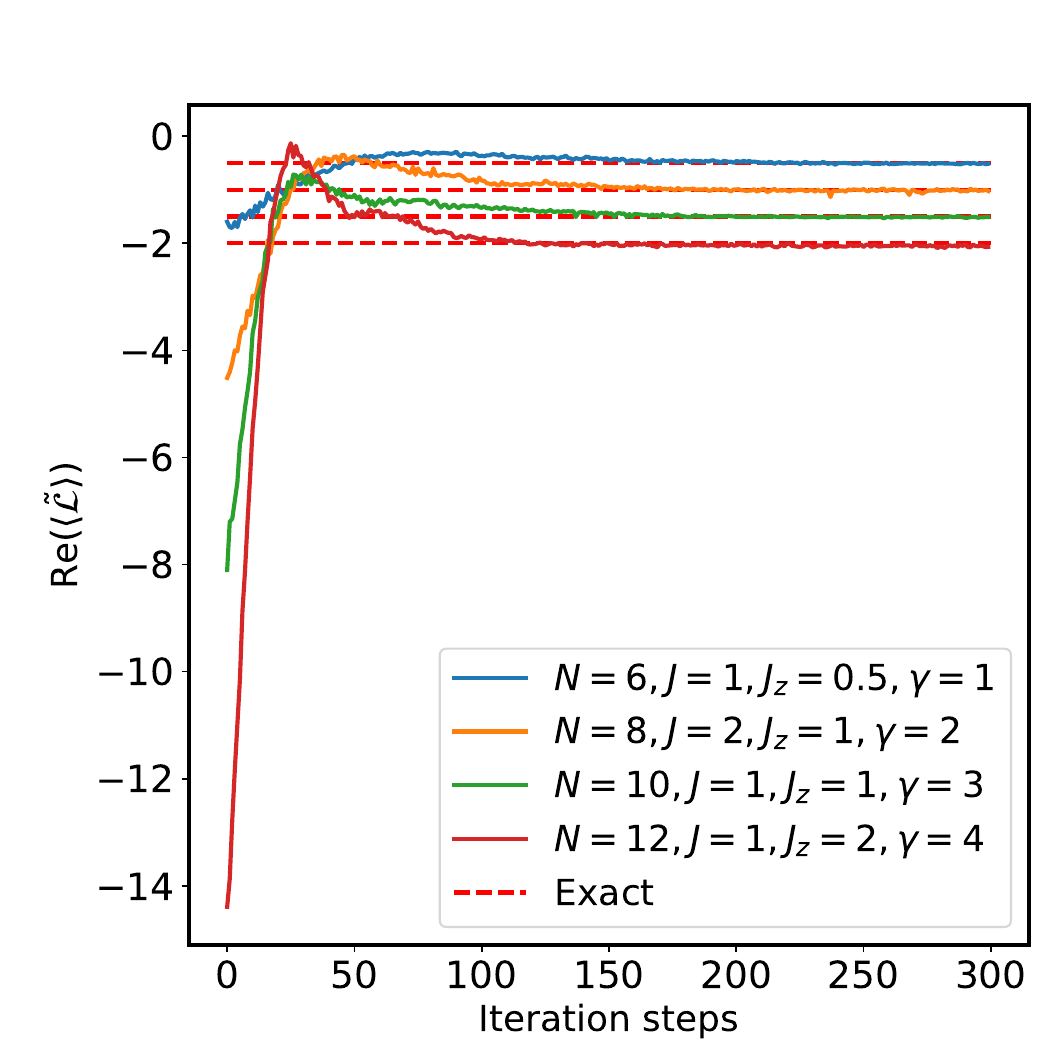}
\caption{The convergence of the Liouvillian gap  for the 1D dissipative long-range XXZ model (See Eqn. \ref{equation: long-range}), obtained by the RBM approach. Different lines correspond to different lattice size $N$ and model parameters $J,J_z,\gamma$. $\alpha=1$. The ancillary $\rho'_0$ used here corresponds to the bi-base state with all spins pointing down.
} \label{figure: SMfig3}
\end{figure}

In Fig. \ref{figure: SMfig3}, we display the convergence behaviour for Liouvillian gap computation of 1D dissipative \textit{long-range} XXZ model obtained by the RBM. We substitute the nearest coupling of the Hamiltonian in the main text Eqn. (6) with the long-range coupling:
\begin{equation}
    H= \sum_{j<k} \left( J^x_{jk} S^x_j S^x_k + J^y_{jk} S^y_j S^y_k + J^z_{jk} S^z_j S^z_k\right) \quad J^x_{jk} = J^y_{jk} = J/ |j-k|^\alpha \quad J^z_{jk} = J_z / |j-k|^\alpha .
    \label{equation: long-range}
\end{equation}
We consider open boundary condition here and the dissipators are the same as those in Eqn. (6) of the main text. Observed from this figure, the real part of the expectation value $\text{Re}(\avr{\tilde{\mathcal{L}}})$ converges quickly to the exact value of the Liouvillian gap, validating the effectiveness of the RBM method to lattice models with long-range interactions.

Finally, in this paper, we mainly discuss the computation of Liouvillian gap for open quantum systems with only one steady state. In order to tackle the cases with multiple steady states by the RBM, the orthogonalization process should be more complicated due to the necessary introduction of left and right eigenstates, which will be left for future explorations.

\end{document}